\pdfoutput=1
\documentclass[a4paper,fleqn]{cas-dc}

\usepackage[numbers]{natbib}
\usepackage{amsmath,amssymb,amsfonts}
\usepackage{algpseudocode}

\usepackage[nolist, withpage]{acronym}
\usepackage{pifont}
\usepackage{tabularx}
\usepackage{multirow}
\usepackage{makeidx}  
\makeindex
\usepackage{pifont}
\usepackage{url}
\usepackage{subfigure}
\usepackage{color}
\usepackage{multirow}
\usepackage{listings}
\usepackage[ruled,boxed, vlined,linesnumbered,resetcount]{algorithm2e}
\usepackage{tikz}
\usepackage{xcolor}
\usepackage{array}
\usepackage{afterpage}
\usepackage{placeins}

\usepackage{hyperref}
\hypersetup{
      colorlinks,%
      citecolor=blue,%
      filecolor=black,%
      linkcolor=black,%
      urlcolor=black
}

\newlength{\figsize}
\newlength{\subfigwidth}
\newlength{\subfiglabelwidth}
\def\tsc#1{\csdef{#1}{\textsc{\lowercase{#1}}\xspace}}
\tsc{WGM}
\tsc{QE}
\tsc{EP}
\tsc{PMS}
\tsc{BEC}
\tsc{DE}

\begin{document}
\let\WriteBookmarks\relax
\def\floatpagepagefraction{1}
\def\textpagefraction{.001}
\shorttitle{Enhanced Dynamic Beamwidth Selection-based THz MAC Protocol for Wireless Data Center Networks}
\shortauthors{M. Absaruddin et~al.}

\title [mode = title]{Enhanced Dynamic Beamwidth Selection-based THz MAC Protocol for Wireless Data Center Networks}                      
\tnotemark[1]

\tnotetext[1]{A preliminary version of this paper has been published in 20th International Conference on Wireless and Mobile Computing, Networking and Communications (WiMob 2024) Paris, France, October 2024 \cite{10770372}.}


\author[knchair]{Muhammad Absaruddin}[orcid=0000-0001-6569-6430]
\cormark[1]
\ead{L00177821@atu.ie}

\author[knchair]{Saim Ghafoor}

\author[rhebo]{Mubashir Husain Rehmani}


\affiliation[knchair]{organization={Department of Computing, Atlantic Technological University},
                    addressline={Donegal}, 
                    country={Ireland}}

\affiliation[rhebo]{organization={Department of Computer Science, Munster Technological University},
                    addressline={Cork}, 
                    country={Ireland}}

\cortext[cor1]{Corresponding author}


\nonumnote{This research was supported by President’s Research Bursary Scholarship Award offered by Atlantic Technological University, Donegal, Ireland.}

\begin{abstract}
Terahertz (THz) wireless communication offers a promising alternative to traditional wired links in data centers (DCs), enabling ultra high data rates, low latency, and greater scalability. However, THz signals suffer from high path loss, necessitating the use of directional antennas (DAs). While DAs enhance signal strength, they introduce challenges such as deafness and synchronisation, typically addressed through receiver initiated MAC protocols. Most existing THz MAC protocols use fixed beamwidths, which results in a key performance trade off: narrow beams improve gain for long range links but reduce throughput for short distances due to increased alignment overhead, while wide beams benefit short links but degrade performance over longer distances. To overcome this limitation, we propose DBS-ADAPT, a dynamic beamwidth selection based MAC protocol that adjusts the antenna beamwidth according to the distance between nodes, maximising throughput without compromising link range. We also introduce an enhanced version, EDBS-ADAPT, which further reduces beamwidth switching and control overhead while preserving throughput gains. Both protocols are evaluated using the NS-3 THz module. Simulation results show that DBS-ADAPT improves average throughput by up to 22\% and reduces delay up to 10\% compared to the baseline ADAPT-3 protocol. EDBS-ADAPT further cuts beamwidth switching overhead by 95\%, making it more efficient for scalable and high performance wireless DC environments.
\end{abstract}


\maketitle

\section{Introduction}
\label{sec:introduction} 
Data center (DC) traffic has been growing exponentially, driving an urgent need for higher bandwidth and more efficient data transmission \cite{geresu2024odrad}. Traditional wired infrastructure and network topologies struggle to keep pace with these demands due to their high deployment and maintenance costs, physical space requirements, and limited flexibility in reconfiguration \cite{han2026wires}. In contrast, wireless communication technologies offer a more flexible, scalable, and cost effective alternative. They also enable dynamic resource allocation and improved adaptability compared to fixed wired links \cite{eckhardt2024channel}, \cite{castro2022thz}.

Several wireless technologies have been explored for DC applications, including millimeter-wave (mmWave) \cite{terzi202160}, optical wireless communication (OWC)\cite{zhang2023photonic}, and Terahertz (THz) communication \cite{eckhardt2024channel}]. Among these, THz communication stands out as a highly promising candidate due to its ability to support ultra high data rates, and wide bandwidth, while offering enhanced physical layer security compared to mmWave. Unlike OWC, THz communication experiences lower beam spreading and penetration losses, and requires less precise alignment, making it more robust to beam misalignment and tracking errors, which are key considerations in dynamic DC environments. Moreover, OWC systems are more vulnerable to environmental challenges such as atmospheric absorption, turbulence, scintillation, and ambient light interference, which can significantly impair performance in practical settings \cite{aboagye2024multi}, \cite{elayan2019terahertz}, \cite{akyildiz2022terahertz}, \cite{alqaraghuli2023road}.

Terahertz wireless links in data centers enable terabit per second data rates, ultra high throughput, and low latency communication, \cite{zhao2025terahertz}, \cite{han2026wires}. They also support flexible, energy efficient topologies that reduce deployment costs \cite{ghafoor2021next}, \cite{ghafoor2020mac}, \cite{absaruddin2023comparative}. In DC environments, THz links can be used for both inter rack between Top of Rack switches and intra rack within a rack communication \cite{cheng2019characterization}. However, THz bands suffer from severe path losses, up to 82 dB at 300 GHz over just 1 meter, with losses increasing at higher frequencies and distances \cite{akyildiz2022terahertz}. These losses can be mitigated using highly directional antennas (DAs) at both transmitter and receiver ends \cite{elayan2019terahertz}, \cite{boulogeorgos2021directional}, \cite{petrov2024wavefront}. Directional antennas are essential in THz communication due to their ability to provide high antenna gain, improve signal to noise ratio, and reduce interference through spatially confined beams \cite{eckhardt2024channel}. These benefits enhance link reliability, spectral efficiency, and energy efficiency while supporting secure and high capacity communication.

While essential for overcoming high path loss, DAs introduce several challenges. These include beam misalignment, directional deafness, and high overhead for beam training and tracking in static as well as dynamic and mobile scenarios \cite{xia2019link}, \cite{ghafoor2020mac}, \cite{akyildiz2022terahertz}. Directional THz links are also highly susceptible to blockage and require precise synchronisation between nodes, complicating medium access control (MAC) and increasing protocol complexity in dense networks. These issues necessitate redesigning link layer protocols to ensure robust and low latency communication. Although rotating directional antennas can help to address some challenges with increasing benefits and usage. They introduced challenges such as frequent beam misalignment, increased overhead of beam tracking and switching, synchronisation difficulties, link stability, beam adjustment for coverage, and beam parameters optimisation. These trade offs require careful protocol design to balance improved coverage and flexibility against increased complexity and energy consumption \cite{akyildiz2016realizing}, \cite{AKYILDIZ201416}, \cite{akyildiz2022terahertz}, \cite{ghafoor2020mac}.

Traditionally, MAC protocol design has concentrated on managing contention among nodes competing for channel access. However, in terahertz communications, the emphasis shifts toward MAC protocols that incorporate scheduling and coordination \cite{ghafoor2020mac}, \cite{absaruddin2023comparative}. In THz communication, we mainly have a transmitter (Tx) and a receiver (Rx) initiated MAC protocols \cite{ghafoor2020mac}. Tx-initiated THz MAC protocols depend on the sender to start communication, which can cause higher collision rates in highly directional environments. In contrast, Rx initiated protocols improve coordination and reduce unnecessary transmissions as transmission attempt is not wasted because the sender only sends data after receiving confirmation from the receiver. This helps to address beam alignment and deafness challenges \cite{ghafoor2020mac}. 

The existing receiver initiated THz MAC protocols \cite{morales2021adapt}, \cite{xia2019link}, \cite{siddiqui2023link}, \cite{siddiqui2023enabling}, \cite{wang2022efficient} predominantly use fixed beamwidth antennas, without accounting for beamwidth dynamics. However, beamwidth selection presents important trade offs: wider beams offer higher throughput and lower delay but at the expense of reduced range, whereas narrower beams provide longer range but result in lower throughput and higher delay \cite{absaruddin2024beamwidth}. Using a fixed beamwidth limits the ability of nodes to adapt to varying distances, leading to suboptimal performance, particularly in terms of throughput and latency. To communicate efficiently, nodes should dynamically adjust their beamwidth based on distance; yet, current receiver initiated protocols do not incorporate distance aware beamwidth adaptation. A balanced strategy is needed for beamwidth selection during the initial discovery and transmission stage \cite{absaruddin2024beamwidth}.

To enable dynamic beamwidth selection, we previously proposed a protocol called Dynamic Beamwidth Selection based THz MAC protocol (DBS-ADAPT) \cite{10770372}, designed to enhance the performance of the existing receiver initiated THz MAC protocol, ADAPT. DBS-ADAPT demonstrated improved average throughput and reduced average delay by allowing both client and Access Point (AP) nodes to use wider beamwidths during the data transmission phase \cite{10770372}. Despite its advantages, DBS-ADAPT has certain limitations: it incurs significant overhead due to frequent antenna beamwidth switching and control packets transmission and involves repeated beamwidth selection processes. In this paper, we present an enhanced version of DBS-ADAPT, called EDBS-ADAPT, which significantly reduces beamwidth switching and control overhead while maintaining the throughput and delay benefits of the original protocol. In our earlier work, a bit error rate (BER) of $10^{-6}$ was assumed, which negatively impacted the overall reliability and Quality of Service (QoS) for data centers. Furthermore, the protocol was based on physical layer parameters from the now outdated IEEE 802.15.3e-2017 THz standard, and its performance was not evaluated under realistic, bursty traffic patterns typically observed in DC environments.

\subsection{Contributions}
The main contributions of our work are summarised as follows:
\begin{enumerate}
\item \textbf{Dynamic Beamwidth Selection protocol:}
In this work, we propose a dynamic beamwidth selection protocol for THz based data center networks, referred to as DBS-ADAPT. This protocol is designed to enhance link performance by adaptively adjusting the beamwidth according to the distance between nodes. The foundational version of this protocol was initially introduced in \cite{10770372}, where its core mechanisms and preliminary results were discussed.

\item \textbf{Proposal of EDBS-ADAPT for Improved Efficiency:}
To address the limitations identified in DBS-ADAPT, we propose an Enhanced Dynamic Beamwidth Selection based THz MAC protocol (EDBS-ADAPT). Designed specifically for DC environments, EDBS-ADAPT significantly reduces antenna beamwidth switching and control overhead by eliminating the repetition of the beamwidth selection process, while preserving the throughput and delay advantages of the original protocol.

\item \textbf{Extension of DBS-ADAPT with Realistic Parameters:}
We extend our previously proposed Dynamic Beamwidth Selection based THz MAC protocol \cite{10770372} by incorporating more realistic system level parameters. Specifically, we adopt a BER of $10^{-12}$, aligning with ITU technical specifications, ETSI recommendations, and IEEE standards for THz data center requirements \cite{etsi}, \cite{itudcn}, to ensure ultra reliable and efficient communication where data integrity is critical.

\item \textbf{Adoption of the Latest Standard and Modulation Schemes:}
The performance of DBS-ADAPT is further evaluated using physical layer parameters based on the most recent IEEE 802.15.3-2023 THz standard. This includes the integration of higher order modulation schemes such as 16-APSK and 32-APSK, allowing for improved data rates and spectral efficiency.

\item \textbf{Scalability Analysis:} 
The scalability analysis is conducted for the proposed protocols by evaluating with a larger data center topology of twenty five nodes.

\item \textbf{Evaluation with Realistic Traffic Models:}
The protocol is assessed under a Poisson traffic model as well as a realistic data center traffic pattern characterised by bursty traffic. This ensures a more accurate representation of actual DC network conditions and validates protocol performance under practical load variations.
\end{enumerate}

The structure of the paper is as follows: Section 2 reviews related work on THz MAC protocols. Section 3 introduces the system preliminaries. Section 4 describes the DBS-ADAPT protocol and its operational details. Section 5 presents the EDBS-ADAPT protocol, including its algorithmic implementation and design. Section 6 outlines the simulation setup, including parameters, implementation details, and the performance metrics used for evaluation. Section 7 discusses the simulation results, comparing DBS-ADAPT and EDBS-ADAPT with the ADAPT-3 THz MAC protocol. Finally, Section 8 concludes the paper and presents the future work.

\section{Related Works}\label{sec:related_work} 

{

\begin{table*}[ht]
    \small
    \centering
    \setlength{\tabcolsep}{2pt}
    \renewcommand{\arraystretch}{1.2}

    \begin{tabularx}{\textwidth}
    {|p{0.5cm}|
      p{1.5cm}|
      p{1.5cm}|
     p{1cm}|
      p{3.2cm}|
     p{1.8cm}|
      p{3cm}|
    p{3.6cm}|}
        \hline
        
        Ref  & Tx/Rx initiated & Beamwidth Management & Distance aware & Technique/Features & Channel Access Method & Strength & Limitations \\ \hline
        \cite{lin2019pulse}  & Tx initiated & Adaptive  & No & Beam switching, Energy control, TDMA scheduling & TDMA & Energy efficient, beam flexibility & Not suitable for bursty DC traffic due to fixed TDMA slots, synchronisation issues \\ \hline
        \cite{yao2016tab}   & Tx initiated & Adaptive  & Yes & Assisted beamforming; optimal beamwidth calculated from signal quality and distance & Dual radio & Improved beam alignment, throughput & \multirow{3}{3.8cm}{Increased hardware cost, coordination complexity, high antennas switching delays, large message overhead led to lower transmission speed and increase in latency. Present chip architecture and the standards not support coordination between two radios} \\ \cline{1-7}
        \cite{yao2019multi}   & Tx initiated & Fixed beamwidth  & Yes & Distance based power allocation & Dual Radio and CSMA & Optimises power consumption & \\ \cline{1-7}
        \cite{zhang2021dual}  & Tx initiated & Adaptive beamwidth & No & Beam parameters adjusted based on exchanged beamwidth information & Dual radio and CSMA & Enhanced throughput, dynamic adaptability & \\ \hline
        \cite{morales2021adapt}  & Rx initiated & Fixed beamwidth & No & Receiver initiated alignment, synchronisation improvements & CSMA & Overcomes deafness and synchronisation issues effectively & Fixed beamwidth: Wider beams reduce range; narrow beams limit throughput. \\ \hline
        \cite{siddiqui2023link}  & Rx initiated & Fixed beamwidth & No & Enhanced node rotation capability, extended link discovery & CSMA & Increased flexibility, improved client node alignment & Still limited by fixed beamwidth issues (range throughput trade off) \\ \hline
        \cite{wang2022efficient}  & Rx initiated & Fixed beamwidth & No & Limited data center specific optimizations & CSMA/TDMA & QoS support & Insufficiently detailed MAC attributes, fixed beamwidth limitations remain \\ \hline
    \end{tabularx}

    \caption{THz MAC protocols related works with techniques, strength and limitations.}
    \label{tab:ref_and_merge}
\end{table*}

}

In this section, we review related work on THz MAC protocols, with a particular focus on dynamic beamwidth selection strategies for data center networks. Existing THz MAC protocols are primarily categorised into Tx-initiated and Rx-initiated approaches \cite{ghafoor2020mac}. In Tx-initiated protocols, the transmitter node initiates communication either by directly sending the data packet in a 0-way handshake or by first sending a control packet e.g., Request to Send, (RTS) in a 2-way handshake. Conversely, Rx-initiated protocols involve the receiver node initiating the communication by broadcasting a control packet indicating its readiness to receive data. The transmitter then responds by sending the data packet. This mechanism is particularly relevant in centralised network architectures typical of data centers, where the receiver acts as an AP and the transmitters are client nodes \cite{ghafoor2020mac}.

Several transmitter initiated THz MAC protocols have been proposed that employ dual radio designs or assisted beamforming to address challenges such as beam alignment and deafness in directional communication. TAB-MAC \cite{yao2016tab} introduces an assisted beamforming MAC protocol that uses a lower frequency control plane to aid beam alignment in the THz band. While it improves beamforming accuracy and reduces setup delay, it relies on dual radio hardware, which is not supported in current chip architectures and adds complexity \cite{d20246g}. MBPA-MAC \cite{yao2019multi} proposes a multi beam, on demand power allocation MAC protocol for MIMO based THz networks. It enhances spatial reuse and energy efficiency through adaptive power control across beams, but it inherits similar limitations due to its dependence on dual radios. DRA-MAC \cite{zhang2021dual} presents a Dual Radio Assisted MAC protocol tailored for distributed THz networks, aiming to support robust MAC operations via cross band communication. However, hardware cost, coordination challenges between radios, and antenna switching delays remain major concerns, especially for high performance data center environments \cite{ghafoor2020mac}. RS-MAC \cite{zhai2023dual} enhances link reliability using an intelligent reflecting surface (IRS) based relay selection mechanism in a dual channel MAC design, but it incurs high control overhead and added latency due to reliance on a separate low frequency channel, limiting its scalability. These protocols offer innovative solutions to key MAC layer issues but face limitations such as hardware incompatibility, increased cost, and higher latency, making them less suitable for ultra fast, low latency data center applications.
 
Several Rx-initiated THz MAC protocols (\cite{morales2021adapt}, \cite{xia2019link}, \cite{siddiqui2023link}, \cite{siddiqui2023enabling}, \cite{wang2022efficient}) have been proposed for centralised networks, where the receiver (AP) starts communication by broadcasting a control packet, prompting the client transmitter to send data. In \cite{xia2019link}, a receiver initiated MAC and link layer synchronisation protocol for THz networks to mitigate deafness and improve timing coordination is proposed. The protocol assumes fixed beamwidths and ideal beam alignment, lacking adaptability for varying link distances and spatial conditions in data centers. In \cite{morales2021adapt}, the ADAPT protocol is introduced, which is an adaptive directional antenna MAC protocol that dynamically manages directional scheduling in THz links. ADAPT does not support dynamic beamwidth adjustment, which can limit performance in dense environments where adaptive spatial coverage is critical. In \cite{siddiqui2023link}, an extension of the NS-3 simulator with THz link discovery capabilities, enabling simulation of Rx-initiated MAC behaviour, is presented.  Focused on simulation infrastructure; does not propose or evaluate new MAC strategies or beam adaptation mechanisms. \cite{siddiqui2023enabling} further enhances NS3 Terasim module to enable TCP communication over THz links, addressing transport layer challenges. Concentrates on transport layer support without incorporating adaptive MAC layer techniques or dynamic beamwidth considerations. In \cite{wang2022efficient}, efficient synchronous MAC protocols for THz wireless data centers are proposed, aiming to reduce latency and improve throughput using coordinated scheduling. While tailored for data centers, the approach does not explicitly address dynamic beamwidth control or physical layer variations such as real time BER adaptation.

ADAPT-1 and ADAPT-3 are receiver initiated THz MAC protocols designed for short range applications like WPAN and WLAN to achieve link layer synchronisation \cite{morales2021adapt}. ADAPT-1 uses a 1-way handshake and is an extension of the LLS-MAC protocol \cite{xia2019link}, while ADAPT-3 uses a 3-way handshake and has been enhanced for node discovery \cite{siddiqui2023link} and downlink communication \cite{siddiqui2023enabling}. Both protocols use fixed beamwidths, where wider beams reduce range and narrower beams limit throughput. Previous studies \cite{absaruddin2023comparative} compared these protocols for wireless data centers, and further analysis \cite{absaruddin2024beamwidth} highlighted the trade offs involved in antenna beamwidth selection. In \cite{lin2019pulse}, a PLBS-MAC, which is a pulse level beam switching scheme with adaptive beamwidth, is presented. However, its contention free TDMA approach struggles to handle the bursty traffic common in data centers. The THz MAC protocols proposed for other applications include TAN-MAC for airborne networks \cite{he2022intelligent}, which is for long range covering a 500 meter distance, and CSMA-IIOT for industrial internet of things \cite{cavallero2023applying}. Dynamic beamwidth mechanism was introduced in our preliminary work \cite{10770372}, a closely related mechanisms was later reported in OBA-MAC \cite{korkmaz2025medium}, and STIN-MAC \cite{he2025adaptive} however for THz drone communication and space communication networks, while our work addresses the DC static scenario.

Recent studies (\cite{xia2019link}, \cite{morales2021adapt}) show that receiver initiated protocols achieve higher throughput and better synchronisation than transmitter initiated ones. This is mainly due to improved coordination, which is crucial when using directional antennas at both ends to address high path loss in THz communication. The limitations of these works include reliance on fixed beamwidths and a lack of dynamic beam adaptation, which reduces efficiency in varying link conditions typical of data centers. Additionally, some focus primarily on simulation or transport layer aspects without fully addressing MAC layer challenges like real time coordination, control overhead, and bursty traffic handling. The shortcomings found in these MAC protocols are mentioned in Table 1.

This work proposes a novel THz MAC protocol for data centers that dynamically selects antenna beamwidth based on the distance between nodes, enhancing communication reliability and efficiency. It introduces the first distance aware, dynamic beamwidth selection algorithm within a receiver initiated THz MAC protocol, ensuring all nodes are discovered and communication links are quickly established. By using a single THz radio, the protocol achieves ultra high throughput and reduced delay, making it well suited for next generation data centers environments.

\section{System Preliminaries}\label{sec:background} 
In this section, the system model with the channel model is presented.

\subsection{System Model} 
We consider a wireless data centers network composed of a single AP and 14 client nodes, each equipped with a single THz transceiver and a steerable directional antenna. As illustrated in Figure 1,, the topology comprises of three rows of racks with five racks per row, totaling 15 racks. Each rack is equipped with a single THz transceiver and a steerable directional antenna (ToR), and the focus of this simulation is on inter rack communication. The horizontal distance between adjacent racks within a row is 1.2 meters, while each rack measures 0.58 meters in width and 0.67 meters in depth. In this topology, a THz access point is centrally located at the ToR of one rack, with the assumption that no client node resides in the same rack as the AP. Nodes are deployed in a top of rack configuration and are static. Each node performs a 360 degree sweep to discover and align with intended communication partners \cite{alghadhban2021f4tele}. 

\begin{figure}
	\centering
	\includegraphics[width=.9\columnwidth]{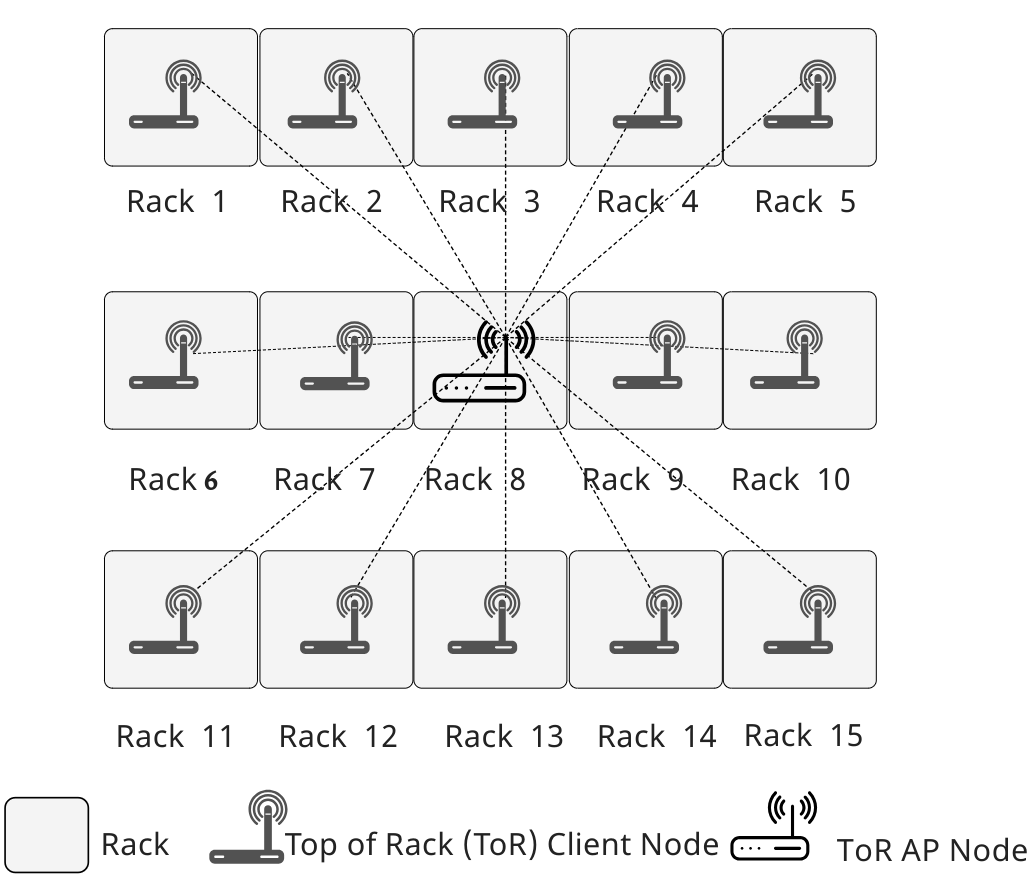}
	\caption{Centralized wireless DC topology with AP and client nodes as top of rack THz nodes.}
	\label{FIG:1}
\end{figure}

\subsubsection{Assumptions} 
To design and evaluate the proposed DBS-ADAPT and EDBS-ADAPT protocols, we base our model on the following realistic and practical assumptions:
\begin{itemize}
\item Directional Antenna Deployment: Both the access point and all client nodes are equipped with steerable directional antennas mounted at the top of rack level. These antennas are capable of beam rotation and dynamic beamwidth adjustment based on node distance.
\item Receiver Initiated Medium Access: The protocol employs a receiver initiated three way handshake using Call to Access (CTA), RTS, and Clear to Send (CTS) frames. This handshake mechanism ensures that only one node is granted transmission access per time slot, eliminating contention and avoiding simultaneous transmissions from multiple clients.
\item Exclusive Time Slot Allocation: After the AP sends a CTS to a selected client, all other client nodes are restricted from transmitting until the time slot ends. This guarantees exclusive use of the wireless channel, thereby preventing beam overlap and co channel interference.
\item Distance Aware Beamwidth Control: Beamwidth is dynamically adapted based on the estimated distance between communicating nodes. This limits unnecessary spatial coverage and further minimises the probability of unintended signal reception.
\item Static Node Placement: All nodes are assumed to be static, which is typical in data centers environments. This assumption simplifies beam tracking and alignment, as node positions do not change over time. It is assumed that the nodes are already aware of their perfect coordinates.
\item Line of Sight (LoS) Availability: The propagation environment is assumed to provide clear LoS paths between racks, which is a reasonable assumption given the structured and obstruction free layout of modern data centers.
\item Interference Negligibility: The combination of directional antennas, exclusive time slot access, and dynamic beamwidth control ensures that interference is minimal, and spatial reuse is efficiently managed.
\item Single Radio Architecture: Each node, including the AP, operates with a single THz radio, eliminating the need for additional hardware and maintaining low system complexity.
\item Synchronised Network Operation: All nodes are time synchronised to a global clock, allowing precise coordination of access slots and handshake operations.
\end{itemize}

\subsection{THz Channel Model} 
In this work, we adopt the THz channel model presented in \cite{jornet2011channel}, which accounts for two key propagation phenomena: free space spreading loss and molecular absorption loss [\cite{jornet2011channel}, \cite{akyildiz2022terahertz}, \cite{elayan2019terahertz}].

The received power is calculated as \cite{jornet2011channel}:
\begin{equation}
Power_{r}(d) = \int_{B} PD_{t}(f) |L_{c}(f,d)|^2 Gain_{t}(f) Gain_{r}(f) df,
\end{equation}

where $L_{c}(f,d)$, the channel frequency response, is determined based on the losses, $Gain_{t}, Gain_{r}$ represent the transmitter and receiver antenna gains, respectively, and $PD_{t}$ denotes the one sided power spectral density of the transmitted signal.

\section{Dynamic Beamwidth Selection based THz MAC Protocol (DBS-ADAPT)}\label{sec:neuralnetworks}
\begin{figure}
	\centering
	\includegraphics[width=.9\columnwidth]{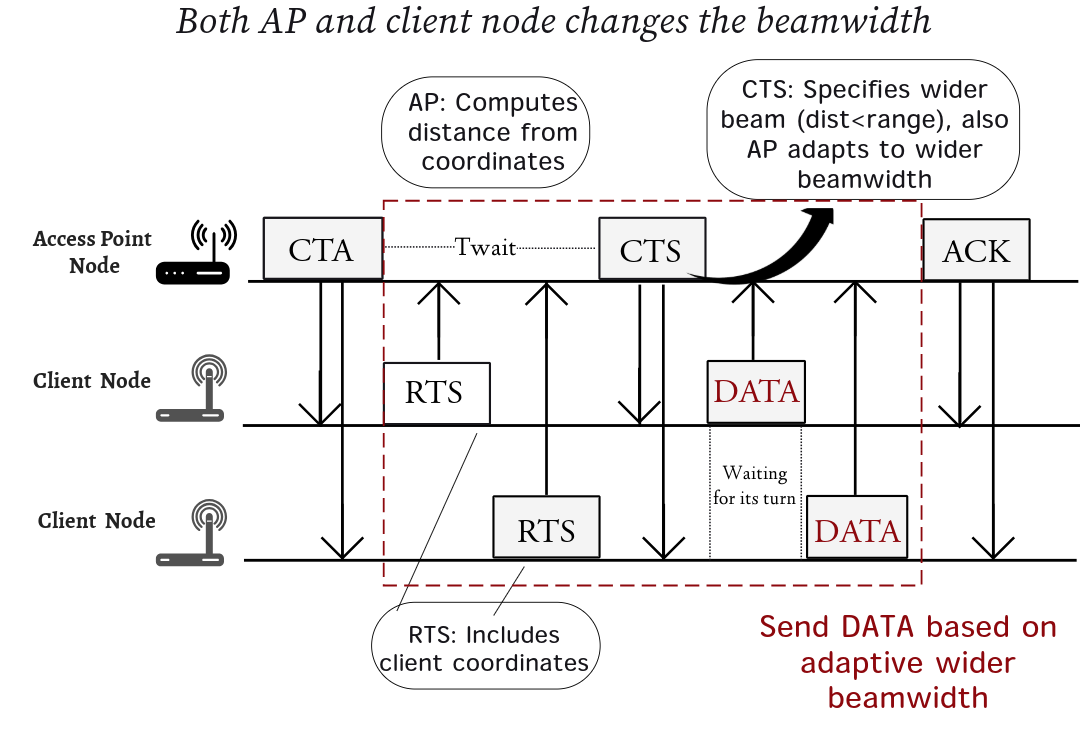}
	\caption{Mechanism of DBS-ADAPT in which both AP and client nodes changes beamwidth during data transmission.}
	\label{FIG:1}
\end{figure}
The DBS-ADAPT \cite{10770372} THz MAC protocol is an enhancement of the ADAPT-3 protocol, specifically designed for wireless data center environments. It is a receiver initiated MAC protocol that overcomes the limitations of fixed beamwidth antennas by dynamically adjusting the antenna beamwidth at both the Access Point and client nodes during the data transmission phase. Initially, narrow beams are employed during link discovery to maximise node detection and alignment accuracy. After successful link establishment and control packet exchange, the AP calculates the distance to the client node and instructs both nodes to adapt their beamwidths dynamically, widening the beam for data transmission if the distance is within range constraints. This distance aware beamwidth adaptation balances communication range and throughput, enabling efficient, high speed data transfer without compromising link reliability.\footnote{We presented a preliminary version of DBS-ADAPT in WiMob 2024 \cite{10770372}, this article introduces a fully extended and refined version.} 

\subsection{Protocol description and flow} 
In DBS-ADAPT, each client node transmits its positional coordinates to the AP during the handshake phase. Using this information, the AP calculates the distance, and both the AP and client dynamically adjust their beamwidths based on that distance. Designed for static wireless data center topologies, DBS-ADAPT accommodates clients positioned at varying distances from the AP, as shown Table 2. The detailed procedure is outlined in Algorithm 1.

The DBS-ADAPT protocol operates in four distinct stages, each described below. An overview of the DBS-ADAPT mechanism is illustrated in Figure 2.

\begin{figure}
	\centering
	\includegraphics[width=.7\columnwidth]{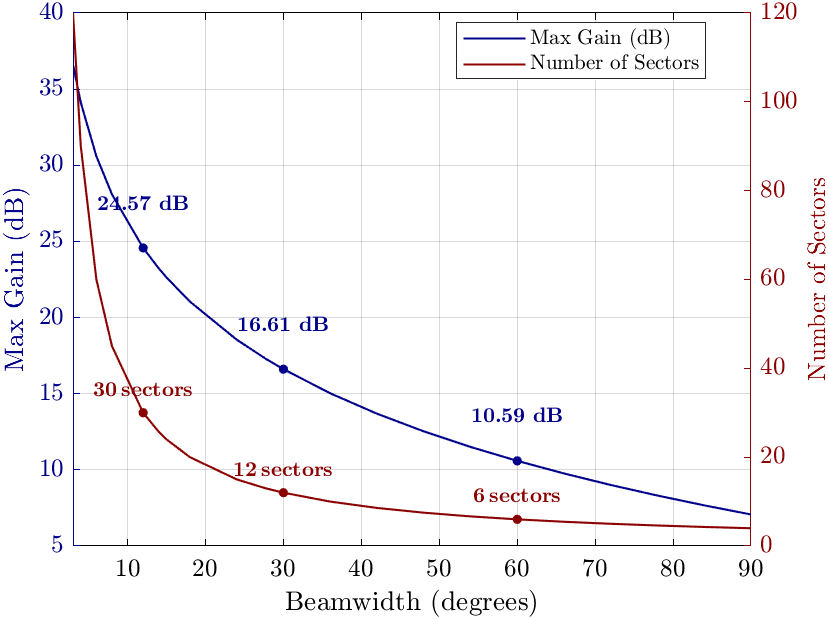}
	\caption{Antenna beamwidth inversely proportional to virtual antenna sectors and gain.}
	\label{FIG:1}
\end{figure}

\begin{enumerate} 
\item \textbf{\textbf{Stage 1: Discovery Stage and Initial Link Establishment}}
Communication begins with the Access Point initiating the discovery process by broadcasting a CTA packet while sweeping its directional antenna. To maximise communication range, the AP initially uses a narrow beamwidth of $6^\circ$, which provides higher antenna gain. At this stage, client nodes are unaware of the AP's location. As the AP sweeps, any client node that receives the CTA packet aligns its antenna toward the direction of the AP using the same fixed narrow beamwidth of $6^\circ$ for accurate and long range detection. This ensures reliable initial link establishment between the AP and client nodes.

\begin{algorithm}
\caption{Algorithm of DBS-ADAPT \cite{10770372}}
\small 

\textbf{Input:} $\Delta \theta$ (Beamwidth), $R(\phi)$ (Range), $\Phi$ - Different beamwidths, 3$^\circ$,6$^\circ$,9$^\circ$,12$^\circ$,15$^\circ$,18$^\circ$,21$^\circ$\ldots\ \\
\textbf{Output:} $\Delta \theta^{\text{dynamic}}$
\begin{algorithmic}[1]
\State Let $\Delta \theta = \{\Phi_1, \Phi_2, \ldots, \Phi_n\}$ 
\State $\Delta \theta_{\text{narrow}}$ $ \leftarrow $ Narrower beamwidth 
\State $\Delta \theta_{\text{wider}}$ $ \leftarrow $ Wider beamwidth 
\State \textbf{(1) CN and AP operations using $\Delta \theta_{\text{narrow}}$}
\State Set initial $\Delta \theta$  = $\min\{\Phi\}$ for CN and AP during discovery 

\State \textit{AP sending the broadcast CTA packet}

\For{each sector in AP coverage area}
\State {Broadcast} {CTA} $\rightarrow$  $\Delta \theta_{\text{narrow}}$ 
 \State       \textbf{end for}

\State \textit{CN Operation of discovering the AP}

\For{each CN}
\State 		\If{\Call{Receive}{CTA}}
        \State \Call{AlignAntenna}{towards AP}
   \State       \textbf{end if}

   \State       \textbf{end for}

\State \textit{CN sending RTS to AP by including its position} 

		\For{each RTS sent by CN}
            \State Include position in RTS packet $\leftarrow$ $(x_n, y_n)$ 
   \State       \textbf{end for}


\State \textbf{(2) AP operation, it calculates distance based on position of CN, and mentions in CTS packet} 

		\For{each RTS received by AP node}
            \State Calculate $d$ between AP and CN.
            \State AP coordinates $\gets$ $(x_{AP}, y_{AP})$
\State $\text{X} \gets x_{AP} - x_n$, $\text{Y} \gets y_{AP} - y_n$
\State $d$ (distance) $\leftarrow \sqrt{\text{X}^2 + \text{Y}^2 }$
			\State Select $\Delta \theta^{\text{dynamic}} = \max\{\Phi\}$ while ensuring  ${d \leq R(\phi)}$
			 \State Set AP $\Delta \theta$  = $\Delta \theta^{\text{dynamic}}$ 
   \State       \textbf{end for}

		\For{each CTS sent by AP node}
            \State Include$\Delta \theta^{\text{dynamic}}$ in CTS.
   \State       \textbf{end for}

\State \textbf{(3) CN $\Delta \theta$ $\rightarrow$  $\Delta \theta_{\text{wider}}$ as mentioned by AP} 

\For{each CTS received by client node}
            \State Extract CTS
			\State Set CN $\Delta \theta$  = $\Delta \theta^{\text{dynamic}}$ for data transmission 
   \State       \textbf{end for}

\end{algorithmic}
\end{algorithm}
A wider beamwidth cannot reach distant nodes effectively due to lower antenna gain, while using an extremely narrow beamwidth, such as $3^\circ$, significantly increases the time required for a full $360^\circ$ sweep. Therefore, a $6^\circ$ beamwidth is chosen as an optimal balance, narrow enough to maintain sufficient gain for reaching the farthest node in the network, yet wide enough to complete discovery efficiently without missing any client nodes. This effect is also illustrated in Figure 3 and Table 2.

\begin{table*}[ht]
    \small 

    \centering
    \setlength{\tabcolsep}{2pt} 
    \renewcommand{\arraystretch}{1.2} 
    \begin{tabularx}{\textwidth}{|p{3cm}|X|X|X|X|X|X|X|X|X|X|X|X|X|X|}
        \hline
        \textbf{Client node ID} & Node 1 & Node 2 & Node 3 & Node 4 & Node 5 & Node 6 & Node 7 & Node 8 & Node 9 & Node 10 & Node 11 & Node 12 & Node 13 & Node 14 \\
        \hline
        \textbf{Distance to AP} & $2.2m$ & $1.95m$ & $1.86m$ & $1.95m$ & $2.2m$ & $1.16m$ & $0.58m$ & $0.58m$ & $1.16m$ & $2.2m$ & $1.95m$ & $1.86m$ & $1.95m$ & $2.2m$ \\
        \hline
        BPSK selected $\Delta \theta$  & $24^\circ$ & $24^\circ$ & $24^\circ$ & $24^\circ$ & $24^\circ$ & $30^\circ$ & $48^\circ$ & $48^\circ$ & $30^\circ$ & $24^\circ$ & $24^\circ$ & $24^\circ$ & $24^\circ$ & $24^\circ$ \\
        \hline

        QPSK selected $\Delta \theta$  & $18^\circ$ & $18^\circ$ & $18^\circ$ & $18^\circ$ & $18^\circ$ & $24^\circ$ & $42^\circ$ & $42^\circ$ & $24^\circ$ & $18^\circ$ & $18^\circ$ & $18^\circ$ & $18^\circ$ & $18^\circ$ \\
        \hline

        8PSK selected $\Delta \theta$  & $12^\circ$ & $12^\circ$ & $12^\circ$ & $12^\circ$ & $12^\circ$ & $18^\circ$ & $30^\circ$ & $30^\circ$ & $18^\circ$ & $12^\circ$ & $12^\circ$ & $12^\circ$ & $12^\circ$ & $12^\circ$ \\
        \hline

        16QAM selected $\Delta \theta$  & $12^\circ$ & $12^\circ$ & $12^\circ$ & $12^\circ$ & $12^\circ$ & $18^\circ$ & $24^\circ$ & $24^\circ$ & $18^\circ$ & $12^\circ$ & $12^\circ$ & $12^\circ$ & $12^\circ$ & $12^\circ$ \\
        \hline

        16APSK selected $\Delta \theta$  & $12^\circ$ & $12^\circ$ & $12^\circ$ & $12^\circ$ & $12^\circ$ & $18^\circ$ & $24^\circ$ & $24^\circ$ & $18^\circ$ & $12^\circ$ & $12^\circ$ & $12^\circ$ & $12^\circ$ & $12^\circ$ \\
        \hline

        32APSK selected $\Delta \theta$  & $12^\circ$ & $12^\circ$ & $12^\circ$ & $12^\circ$ & $12^\circ$ & $12^\circ$ & $18^\circ$ & $18^\circ$ & $12^\circ$ & $12^\circ$ & $12^\circ$ & $12^\circ$ & $12^\circ$ & $12^\circ$ \\
        \hline
    \end{tabularx}
    \caption{Details of beamwidth $\Delta \theta$ selected by each of the client nodes concerning their distance and different modulation and coding schemes for DBS-ADAPT and EDBS-ADAPT THz MAC protocol.}
    \label{tab:example}
\end{table*}

\item \textbf{Stage 2: Exchange of Positional Information and Distance Calculation}
Once a client node aligns its antenna toward the AP and has data to transmit, it initiates the handshake process by sending a RTS packet using a narrow $6^\circ$ beamwidth. The RTS packet includes the client's positional coordinates. Upon receiving the RTS, the AP extracts the coordinates and computes the Euclidean distance to the client node using Equation (5), as shown in Line 21 of Algorithm 1.

\begin{equation} 
d = \sqrt{(x_{\text{AP}} - x_n)^2 + (y_{\text{AP}} - y_n)^2} 
\end{equation} 

where \(d\) is the Distance in meters (m), $(x_{AP}, y_{AP})$ are AP coordinates, and $(x_n, y_n)$ are Client node coordinates. 

This distance is then used to determine the appropriate beamwidth for efficient data transmission.

\item \textbf{Stage 3: Dynamic Beamwidth Selection for AP and Client Nodes} 
Based on the calculated distance, the AP selects an optimal beamwidth that guarantees the communication range exceeds the required distance, as implemented in Line 22 of Algorithm 1. This ensures reliable connectivity while allowing faster antenna rotation, which contributes to improved throughput. The AP then includes the selected beamwidth in the Clear to Send (CTS) packet sent to the client node. Upon receiving the CTS, the client adjusts its antenna beamwidth to match the value specified by the AP. As a result, both the AP and client operate using the same dynamically selected beamwidth. The specific beamwidth values corresponding to various distances and Modulation and Coding Schemes (MCS) are detailed in Table 2.	
 
\item \textbf{Stage 4: Data Transmission with Adjusted Beamwidth} 
For data transmission, the client node uses the newly adjusted beamwidth determined during the handshake phase. Once the AP successfully receives the data packet, it responds with an acknowledgement (ACK) packet. By using dynamically selected beamwidths, both the AP and client can rotate their antennas more quickly, leading to faster link alignment, increased communication opportunities, reduced transmission delay, and overall improved throughput.
\end{enumerate}

The protocol begins with a narrow beamwidth during link discovery to ensure all nodes, regardless of distance, can establish connections while maximising communication range. Then, during control packet exchange, the AP selects an optimal beamwidth based on the client's distance, and the client adjusts accordingly, repeating this process for every data transmission to maintain efficiency.

DBS-ADAPT improves performance primarily through dynamic beamwidth adjustment. In this protocol, the access point determines the appropriate beamwidth based on the distance to the target client node. The client then adjusts its beamwidth according to the value received from the AP. This adjustment occurs for every data packet transmission, leading to frequent beamwidth switching and increased control overhead. While DBS-ADAPT achieves better throughput and delay performance, it does so at the cost of higher overhead. To mitigate this issue, an improved version called Enhanced DBS-ADAPT is proposed.

\section{Enhanced Dynamic Beamwidth Selection based THz MAC Protocol (EDBS-ADAPT)}\label{sec:sba_adapt} 
We propose an enhancement to DBS-ADAPT called Enhanced DBS-ADAPT (EDBS-ADAPT), a receiver initiated THz MAC protocol specifically designed for wireless data centers. EDBS-ADAPT follows the same core mechanism as DBS-ADAPT, with one key difference: in DBS-ADAPT, beamwidth calculation is included in every control packet exchange, causing it to be repeated for each data packet transmission. This results in frequent beamwidth switching and increased overhead. In contrast, EDBS-ADAPT selects the beamwidth only once at the start of communication, maintaining it throughout the session. This significantly reduces beamwidth switching overhead.  Figure 2, also represents the EDBS-ADAPT mechanism the only difference is that both client and AP nodes selects the beamwidth only once, and maintains it throughout the session.

\subsection{Protocol description and flow}
To illustrate the decision making process of the EDBS-ADAPT THz MAC protocol, Algorithm 2 outlines the detailed steps. The protocol's operation is divided into several stages, as described below.

\begin{algorithm}
\caption{Algorithm of EDBS-ADAPT}
\small 
\textbf{Input:} $\Delta \theta$ (Beamwidth), $R(\phi)$ (Range), $\Phi$ - Different $\Delta \theta$ \\
\textbf{Output:} $\Delta \theta^{\text{dynamic}}$
\begin{algorithmic}[1]
\State Let $\Delta \theta = \{\Phi_1, \Phi_2, \ldots, \Phi_n\}$ 
\State $\Delta \theta_{\text{narrow}}$ $ \leftarrow $ Narrower beamwidth, $\Delta \theta_{\text{wider}}$ $ \leftarrow $ Wider beamwidth 
\State \textbf{(1) CN and AP operations using $\Delta \theta_{\text{narrow}}$}
\State Set initial $\Delta \theta$  = $\min\{\Phi\}$ for CN and AP during discovery 
\State \textit{AP sending the broadcast CTA packet}

\For{each sector in AP coverage area}
\State {Broadcast} {CTA} $\rightarrow$  $\Delta \theta_{\text{narrow}}$ 
   \State       \textbf{end for}
\State \textit{CN Operation of discovering the AP}

\For{each CN}
\State 		\If{\Call{Receive}{CTA}}
        \State \Call{AlignAntenna}{towards AP}
   \State       \textbf{end if}

   \State       \textbf{end for}

\State \textit{CN sending RTS to AP by including its position} 

\For{each RTS sent by CN}
\State Include position in RTS packet $\leftarrow$ $(x_n, y_n)$ 
   \State       \textbf{end for}


\State \textbf{(2) Beamwidth Learning and One Time Switching at AP}

		 \For{each RTS received by the AP node}
           \State Calculate $d$ between AP and CN.
          \State AP coordinates $\gets$ $(x_{AP}, y_{AP})$
 \State $\text{X} \gets x_{AP} - x_n$, $\text{Y} \gets y_{AP} - y_n$
 $d$ (distance) $\leftarrow \sqrt{\text{X}^2 + \text{Y}^2 }$
			  \State AP checks if a beamwidth is already stored for this Client:
			  
  \State  If \textbf{no beamwidth is stored} for this Client:
        \begin{itemize}
			 \item AP selects $\Delta \theta^{\text{dynamic}} = \max\{\Phi\}$ while ensuring  ${d \leq R(\phi)}$
			 \State Set AP $\Delta \theta$  = $\Delta \theta^{\text{dynamic}}$ 
			  \item AP stores $\Delta\theta^*_\mathrm{AP,Client} \gets \Delta\theta^*$ for this client in a table.
			          \end{itemize}

   \State       \textbf{end for}
   
  \State     If \textbf{beamwidth already stored} for this Client:
        \begin{itemize}
     \item  AP retrieves $\Delta\theta^*_\mathrm{AP,Client}$ from the table.
        \end{itemize}

 \State  AP includes $\Delta\theta^*_\mathrm{AP,Client}$ in the CTS reply to Client.
 
AP switches beamwidth for this Client only once at first data exchange; subsequent transmissions reuse stored value.

 \State  \textbf{(3) Beamwidth Switching at Client (One Time)}
\begin{itemize}
  \item Upon receiving CTS, reads the $\Delta\theta^*_\mathrm{Client}$ (from CTS).
  \item If the Client has not switched beamwidth before:
    \begin{itemize}
      \item Client switches its antenna to $\Delta\theta^*_\mathrm{Client}$.
      \item Client marks "beamwidth switched" (flag); future data exchanges use the same beamwidth.
    \end{itemize}
  \item If already switched, continue using stored beamwidth.
\end{itemize}

\BlankLine
 \State  \textbf{(4) Data Transmission (No Further Switching)}
\begin{itemize}
  \item Data and ACK packets are exchanged using the $\Delta \theta$  = $\Delta \theta^{\text{dynamic}}$ optimal beamwidth.
  \item No further beamwidth changes are performed for either the AP or the Client node 
\end{itemize}

\end{algorithmic}
\end{algorithm}

\begin{enumerate} 
\item \textbf{Stage 1: Initial link establishment with narrow beams } 
The access point initiates communication using a fixed beamwidth of $6^\circ$, broadcasting CTA packets while rotating through all sectors to signal its readiness to receive data. Since client nodes are initially unaware of the AP's location, each client enters a discovery phase. During this phase, the client uses the same fixed beamwidth and performs exhaustive scanning by rotating its antenna across all sectors, pausing in each one to listen for a CTA packet. When a client detects a CTA packet in a particular sector, it stops scanning, aligns its antenna in that direction, and transitions to the communication phase to begin data exchange with the AP.

\item \textbf{Stage 2: RTS transmission and position exchange} 
The client node generates an RTS packet that includes its position as coordinates $(x_{AP}, y_{AP})$. Upon receiving the RTS packet, the AP extracts the client's coordinates and calculates the distance to the client node.

\item \textbf{Stage 3: AP learning and beamwidth assignment} 
The AP consults its beamwidth mapping table to determine whether an optimal beamwidth for the client has already been established:
\begin{itemize}
  \item   If not previously learned: The AP selects an appropriate beamwidth based on the client's distance and link quality, then stores this information in the mapping table.
  \item   If already learned: The AP retrieves the previously stored beamwidth value for the client.
\end{itemize}

The AP then sends a Clear To Send packet to the client, including the assigned beamwidth value.

\item \textbf{Stage 4: One time beamwidth adjustment} 
Upon receiving the CTS packet, the client extracts the assigned dynamic beamwidth and immediately adjusts its antenna to this value for all future data transmissions to the AP. It also sets a beamwidth adjustment flag to prevent further changes during the session. Similarly, before receiving the first DATA packet from the client, the AP switches its antenna to the selected dynamic beamwidth for that specific client and sets a corresponding per client adjustment flag. The dynamic beamwidth values selected based on distance and various modulation and coding schemes are detailed in Table 2. 
\end{enumerate}

\section{Simulation details} 
\label{sec:concept}  
In this section, we describe our simulation setup, parameters, implementation details, and performance metrics. The simulations are conducted using Terasim, an open source THz network simulator built as an extension of NS3 \cite{hossain2018terasim}. The specific simulation parameters are listed in Table 3. Packet arrivals are modelled using a Poisson process \cite{morales2021adapt}, chosen for its effectiveness in representing random traffic patterns. A log-normal traffic model is also used for bursty traffic. All simulation results presented in Section 7 are reported with a 95\% confidence interval (CI).

\begin{table}[ht] 
    \small 

\centering 

\begin{tabularx}{\columnwidth}{|X|X|X|} 

\hline 

\textbf{Description} & \textbf{Value} \\ 

\hline 

Simulator & Terasim - NS3 \cite{hossain2018terasim} \\  

\hline 

Technology  & IEEE 802.15.3-2023 \cite{10443750} \\  

\hline 

Frequency Range  & 252.72 321.84 GHz \cite{10443750} \\  

\hline 

Central Frequency & 287.28 GHz \cite{10443750} \\ 

\hline 

Channel bandwidth & 69.12 GHz \cite{10443750} \\ 

\hline 

Transmission power  & 10dBm \cite{itudcn} \\  

\hline 

Bit error rate (BER)  & $10^{-12}$ \cite{etsi}, \cite{itudcn}, \cite{ghafoor2020mac} \\  

\hline 

Simulation time for each run &  0.01 seconds \\ 

\hline 

Simulation Runs &  20 \\ 

\hline 

Noise figure & 7dB \\ 

\hline 

Data Packet Size & 65 kB \\ 

\hline 

Retry limit & 5 \\ 

\hline 

Traffic Model & Poisson \cite{morales2021adapt}, Log Normal  \\ 

\hline 

Modulation and Coding Schemes & BPSK, QPSK, 8PSK, 16QAM, 16APSK, and 32APSK \cite{10443750}. \\ 

\hline 

 \end{tabularx} 
\caption{Simulation Parameters\label{tab:table1}} 
\end{table} 

%
\subsection{Simulation Parameters} 
The performance evaluation uses various modulation and coding schemes defined in the recent THz standard IEEE 802.15.3-2023. These include Binary Phase Shift Keying (BPSK), Quadrature Phase Shift Keying (QPSK), 8-Phase Shift Keying (8-PSK), 16-Quadrature Amplitude Modulation (16-QAM), 16-Amplitude and Phase Shift Keying (16-APSK), and 32-Amplitude and Phase Shift Keying (32-APSK) \cite{10443750}. For data center applications, a BER of $10^{-12}$  is adopted, as recommended by the IEEE standard, ITU technical parameters, and ETSI guidelines \cite{itudcn}, \cite{etsi} \cite{ghafoor2020mac} \cite{absar2026issc}. Additionally, the performance is evaluated by varying interarrival times to analyse how the protocol behaves under different traffic loads, reflecting the high variability typical of data center traffic \cite{hu2024load}. The interarrival time is calculated as described in Equation 6.

\begin{equation} 
P = \frac{T}{I} 
\end{equation} 
where \(P\) is number of data packets, \(T\) is the simulation duration in seconds, and \(I\) is the inter-arrival time in seconds. 

Another important parameter in the evaluation is the antenna turning speed, which depends on the beamwidth  and the time required to cover a sector. The turning speed refers to how quickly the antenna completes a full $360^\circ$ rotation. This speed influences how fast the antenna can sweep its coverage area, directly impacting throughput and delay. The antenna turning speed increases proportionally with the beamwidth. In other words, antennas with wider beamwidths rotate faster than those with narrower beamwidths, allowing nodes to complete rotations more quickly.

The antenna turning speed is calculated using the following expression:
\begin{equation}
\omega = \frac{1}{T_\text{circle}}
\end{equation}

where $\omega$ is the Antenna turning speed measured in rotations per second (degrees/sec), $T_\text{circle}$ is the total time for one complete antenna rotation, measured in seconds. The total rotation time  $T_\text{circle}$ can be calculated as:

\begin{equation}
T_\text{circle} = T_\text{sector} \cdot N_\text{sector}
\end{equation}
where $T_\text{sector}$ is calculated as the sum of control packets duration, propagation time, SIFS, maximum backoff time, data transmission time, and $N_\text{sector}$ is calculated as $ \frac{360}{\text{beamwidth}} $.

\begin{figure}
	\centering
	\includegraphics[width=.9\columnwidth]{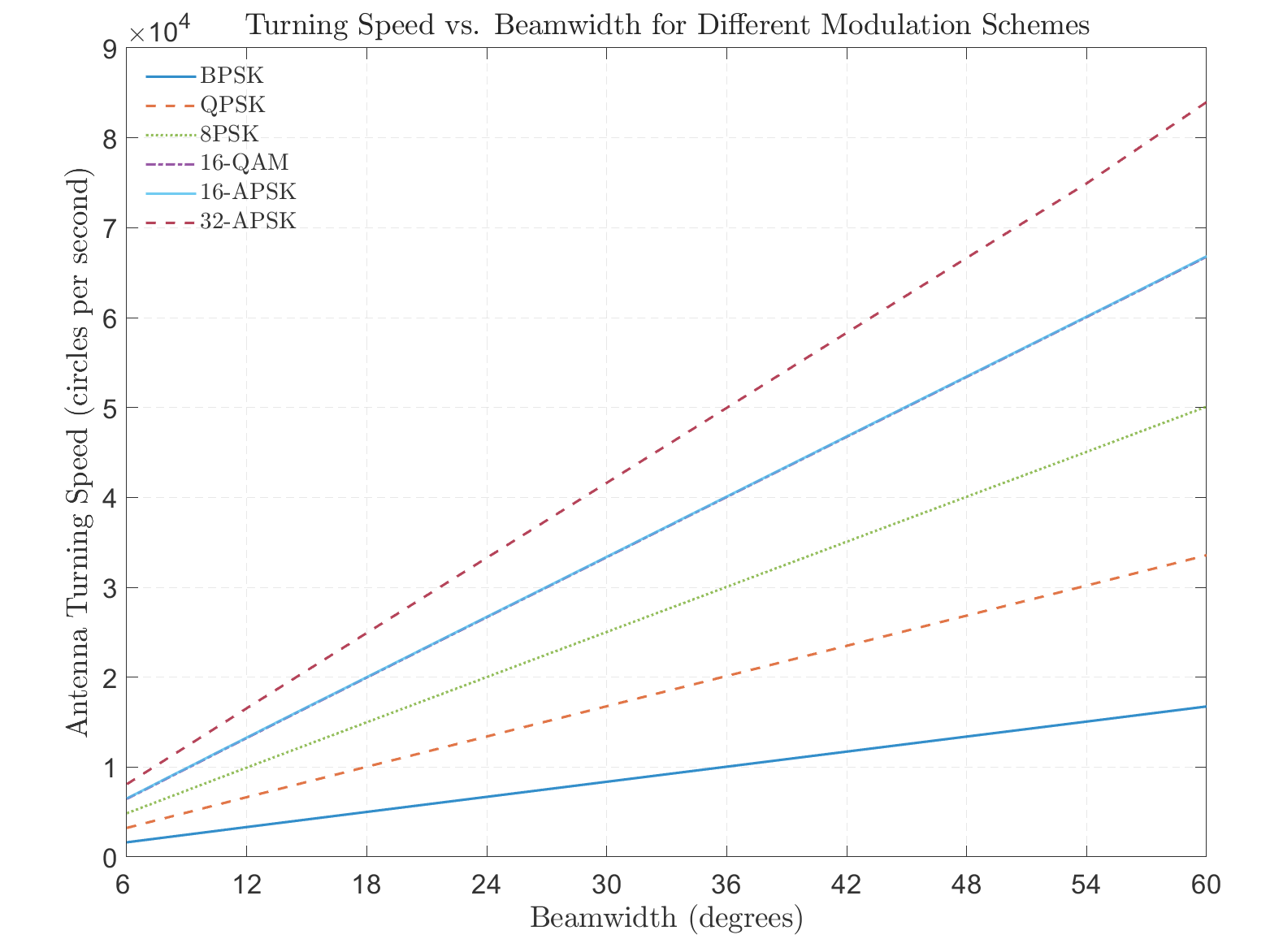}
	\caption{Centralized wireless DC topology with AP and client nodes as ToR THz nodes.}
	\label{FIG:1}
\end{figure}

\subsection{Implementation Details} 
For DBS-ADAPT, the Terasim simulator was modified to include client node position information in the RTS control packet header. Specifically, additional fields were added to the RTS packet to store the client's $x-y$ coordinates. This allows the AP to calculate the distance to the client node. Similarly, the CTS packet was updated with a new field to carry the beamwidth information, which is sent back to the client. A new mechanism was implemented at the MAC layer to enable dynamic beamwidth selection during data transmission. Additionally, the antenna module was modified to update the beamwidth, recalculate antenna gain whenever the Tx or Rx changes beamwidth, and adjust the antenna turning speed accordingly.

For EDBS-ADAPT, we modified the NS3 THz MAC modules for both AP and client nodes to implement a low overhead beamwidth switching mechanism. Unlike previous approaches that adjusted beamwidth frequently, our design ensures that both AP and clients switch their antenna beamwidth only once per communication session. Specifically, after receiving the CTS packet with the optimal beamwidth, the client sets its antenna accordingly and maintains this setting throughout the session. This one time adjustment is enforced in the client's MAC module, significantly reducing protocol overhead.

On the AP side, the protocol is optimised by maintaining a mapping between each client's address and its learned optimal beamwidth. When an RTS packet is received, the AP either retrieves the stored beamwidth for that client or calculates a new one based on the client's position and distance. This beamwidth is sent to the client via the CTS packet. After the first data transmission, the AP saves the beamwidth for future use, avoiding repeated adjustments. A per client flag ensures that the AP switches beamwidth only once per communication session.

\subsection{Performance Metrics} 
The performance of our proposed DBS-ADAPT and EDBS-ADAPT THz MAC protocols is evaluated using three key metrics: average throughput, average delay, and beamwidth switching overhead, as described below.

\begin{enumerate}  
\item \textbf{Average throughput:} This metric measures the effectiveness of data delivery. It represents the rate at which data packets are successfully transmitted by each client node. Average throughput is calculated by averaging the throughput of all individual client nodes, reflecting the overall efficiency of data transmission. It is typically measured in bits per second (bps) \cite{absaruddin2024beamwidth}.

The average throughput is calculated as follows \cite{absaruddin2024beamwidth}: 
\begin{equation} 
T_{\text{avg}} = \frac{1}{N} \sum_{n=1}^{N} \frac{L_n}{t_n} 
\end{equation} 
Where \(T_{\text{avg}}\) is the average throughput (in bps or Gbps), \(N\) is the total number of client nodes, \(L_n\) is the total size of packets transmitted by node \(n\) (in bits), and \(t_n\) is the total transmission time for node \(n\) (in seconds).
 \item \textbf{Average delay:} This metric is measured from the moment a packet is queued for transmission until it is successfully received by the destination node. It accounts for transmission time, queuing delays, MAC protocol processing time, and propagation delay. Average delay is a key indicator of the protocol's efficiency and QoS in THz communication systems \cite{terapod}. 

The average delay is calculated as follows \cite{terapod}: 
\begin{equation} 
D_{\text{avg}} = \frac{1}{N} \sum_{i=1}^{N} \left(T_{\text{end},i} - T_{\text{start},i}\right) 
\end{equation} 
where \(D_{\text{avg}}\) is the average delay measured in \textit{milliseconds (ms)}, $T_{\text{end},i}$ is the time when the transmission of the $i$-th packet is completed.
 \item \textbf{Beamwidth switching overhead:} This metric represents the total number of times a node changes its beamwidth during communication.
\end{enumerate} 

\section{Simulation Results}\label{sec:periodicity}

\begin{figure}[h]
    \centering
    \subfigure[Average throughput with BPSK, QPSK, and 8PSK MCS.]{\label{fig:varying_thr_tra_poisson} \includegraphics[width=0.38\textwidth]{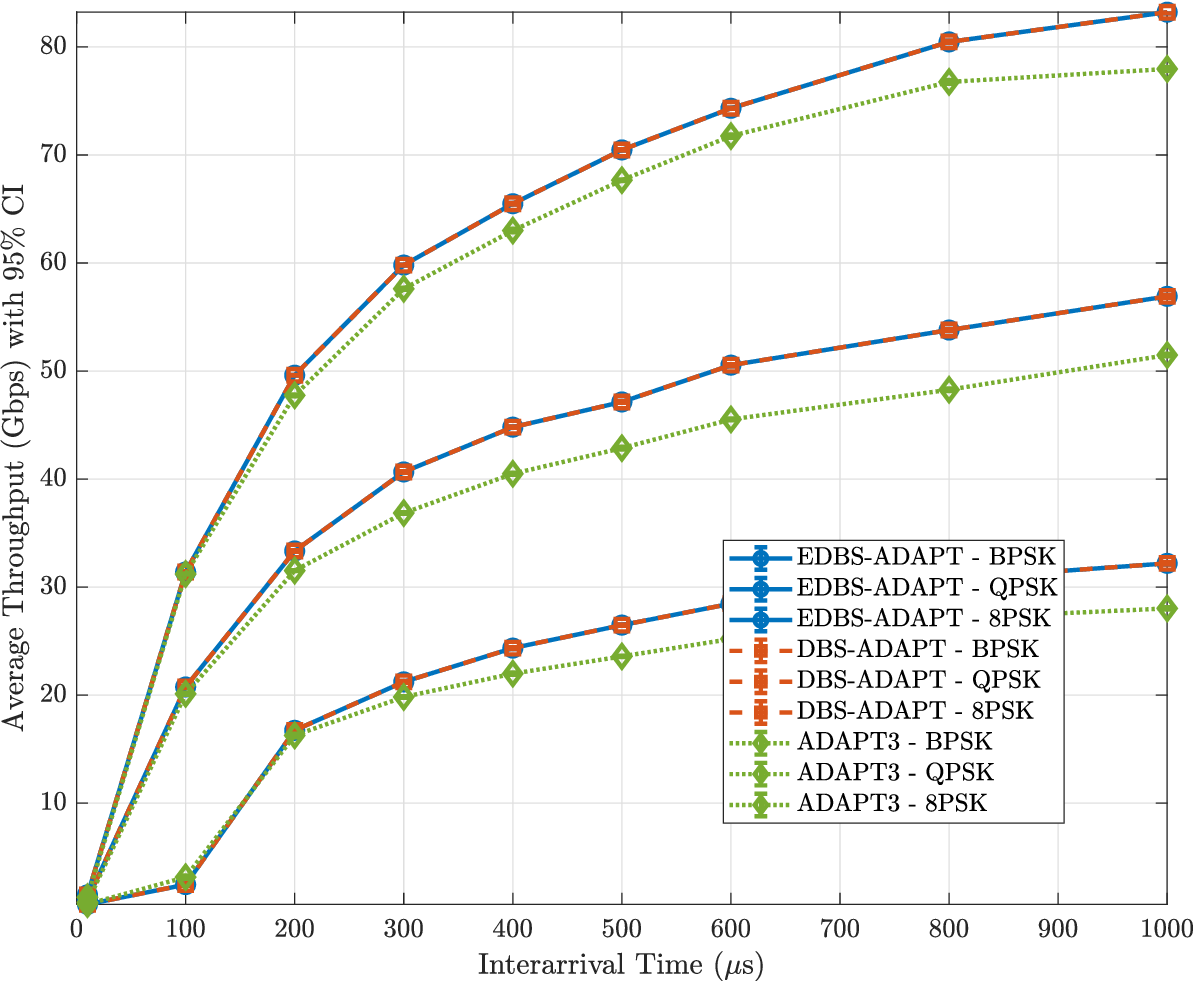}}
    \hfill
    \subfigure[Average delay with BPSK, QPSK, and 8PSK MCS.]{\label{fig:varying_ad_tra_poisson} \includegraphics[width=0.38\textwidth]{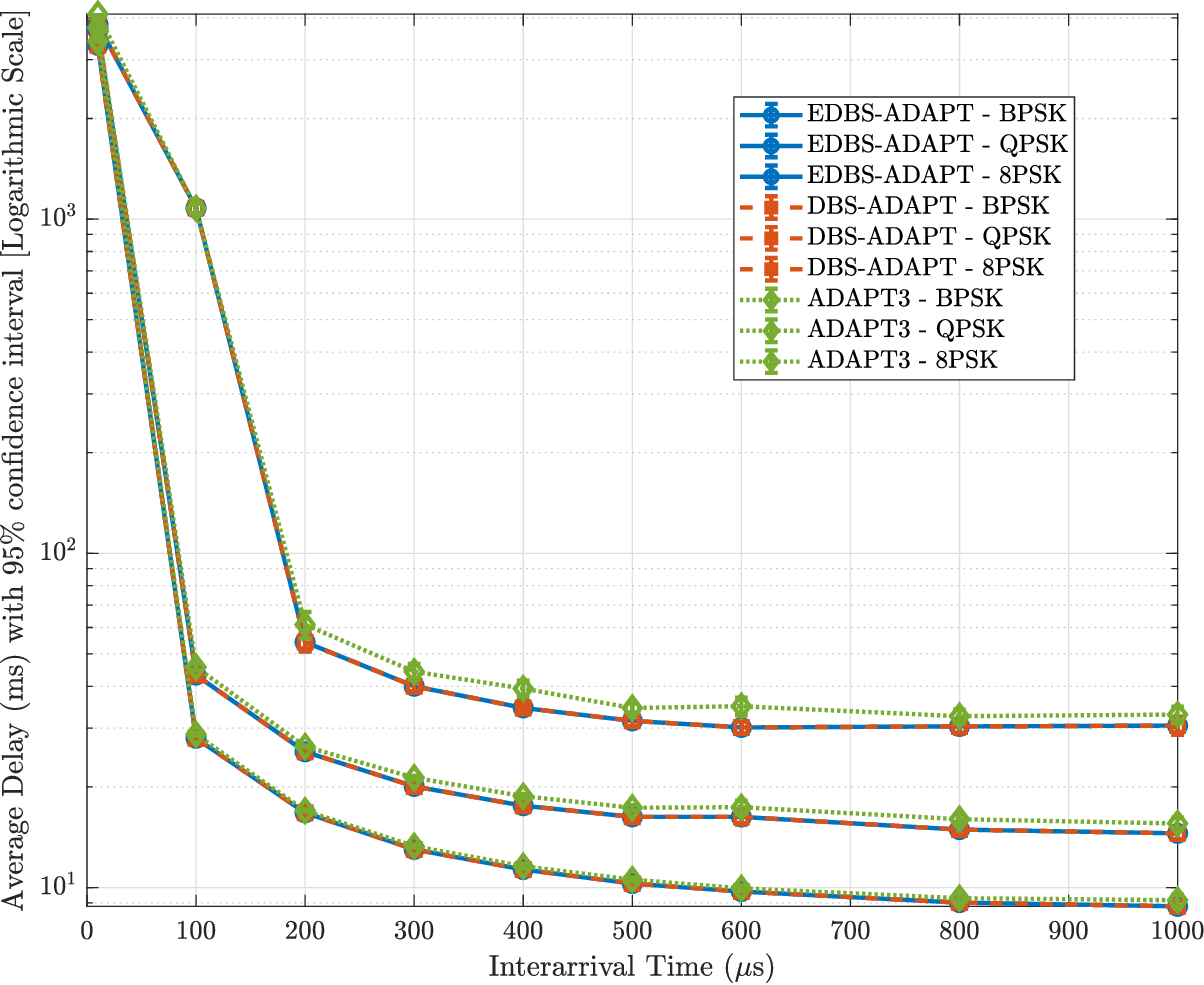}}
    \caption{EDBS-ADAPT 14 nodes case with varying inter-arrival time.}
    \label{fig:17}
\end{figure}

In this section, we present the performance results of DBS-ADAPT and EDBS-ADAPT compared to the baseline THz MAC protocol, ADAPT-3 \cite{morales2021adapt}, under both Poisson and bursty traffic models.

\begin{table}[h!]
\scriptsize 
\centering
\begin{tabular}{|p{0.8cm}|p{0.4cm}|p{0.4cm}|p{0.4cm}|p{0.4cm}|p{0.4cm}|p{0.4cm}|p{0.4cm}|p{0.4cm}|p{0.4cm}|}
\hline
Interarrival Time ($\mu\text{s}$) & 10 & 100 & 200 & 300 & 400 & 500 & 600 & 800 & 1000 \\ \hline
No of data packets   & 1000   & 100   & 50   & 33   & 25   & 20   & 16   & 12   & 10   \\ \hline
\end{tabular}
\caption{Number of packets sent against each selected inter-arrival time.}
\label{tab:example_table}
\end{table}

\subsection{Simulation Results with Poisson Traffic Model}\label{sec:periodicity8}
DBS-ADAPT is evaluated against ADAPT-3 \cite{morales2021adapt} using modulation and coding schemes (Section 5.3) and a BER of $10^{-12}$, as specified in the latest IEEE 802.15.3-2023 standard. This comparison demonstrates the effectiveness of DBS-ADAPT under realistic THz communication parameters for data centers. EDBS-ADAPT is further compared against both DBS-ADAPT and ADAPT-3. The proposed protocols are analysed based on varying interarrival times, number of nodes, and modulation schemes. In ADAPT-3, a fixed beamwidth of 6 degrees is used for both the discovery and data transmission phases. This beamwidth was chosen based on evaluations within a data center topology, offering a balance between range and antenna gain to mitigate high path loss and ensure stable performance.

\subsubsection{EDBS-ADAPT, and DBS-ADAPT with varying inter-arrival times and modulation schemes}
EDBS-ADAPT and DBS-ADAPT are evaluated under varying interarrival times and modulation schemes to assess their performance under different traffic loads. Interarrival times, ranging from 10 to 1000 $\mu\text{s}$, follow an exponential distribution and represent the time between the arrivals of two consecutive packets at a node. This variation helps capture the random and highly variable nature of data center traffic \cite{hu2024load}. For analysis, traffic is categorised into high and low traffic scenarios: high traffic corresponds to interarrival times between 1-299 $\mu\text{s}$, while low traffic ranges from 300-1000 $\mu\text{s}$. Table 4 presents the number of data packets sent by each client based on these interarrival times, and is calculated using the equation 3.

In addition to traffic variability, performance is evaluated across six modulation and coding schemes (MCS): BPSK, QPSK, 8-PSK, 16-QAM, 16-APSK, and 32-APSK. These MCSs are essential for enabling efficient and flexible communication in THz networks. Their varying encoding capabilities directly influence key performance metrics such as average throughput and delay. Higher-order modulation schemes, such as 16-QAM, 16-APSK, and 32-APSK, are particularly important for achieving the ultra-high throughput and low latency required in data center environments.

\paragraph{\textbf{Average Throughput and delay with varying inter-arrival times and modulation schemes:}}
Figures 5(a), 5(b), 6(a), and 6(b) present the average throughput and delay results for EDBS-ADAPT, DBS-ADAPT, and ADAPT-3 under a fixed 14-node setup with varying interarrival times. In scenarios using the low modulation scheme BPSK, EDBS-ADAPT and DBS-ADAPT consistently outperforms ADAPT-3 in terms of average throughput. For instance, under low traffic conditions with an interarrival time of 1000 $\mu\text{s}$, EDBS-ADAPT and DBS-ADAPT achieves an average throughput of 32 Gbps, compared to 28 Gbps for ADAPT-3, an improvement of approximately 15\%. When the interarrival time is reduced to 500 $\mu\text{s}$, throughput for EDBS-ADAPT and DBS-ADAPT is 26 Gbps versus 23 Gbps for ADAPT-3, marking a 13\% gain.

\begin{figure}[h]
    \centering
    \subfigure[Average throughput with 16QAM, 16APSK, and 32APSK MCS.]{\label{fig:varying_thr_tra_poisson_2} \includegraphics[width=0.38\textwidth]{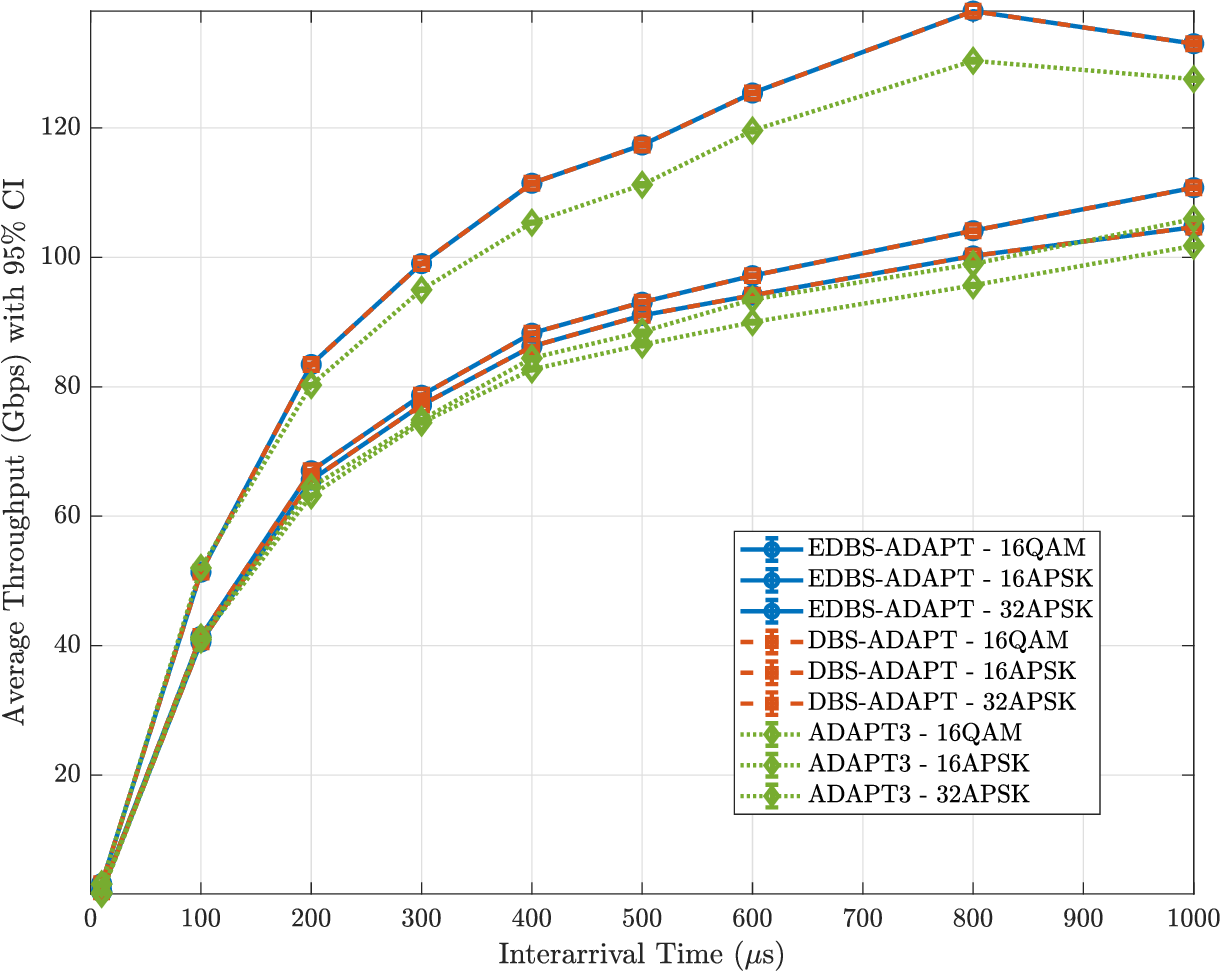}}
    \hfill
    \subfigure[Average delay OF DBS-ADAPT with 16QAM, 16APSK, and 32APSK MCS.]{\label{fig:varying_ad_tra_poisson_2} \includegraphics[width=0.38\textwidth]{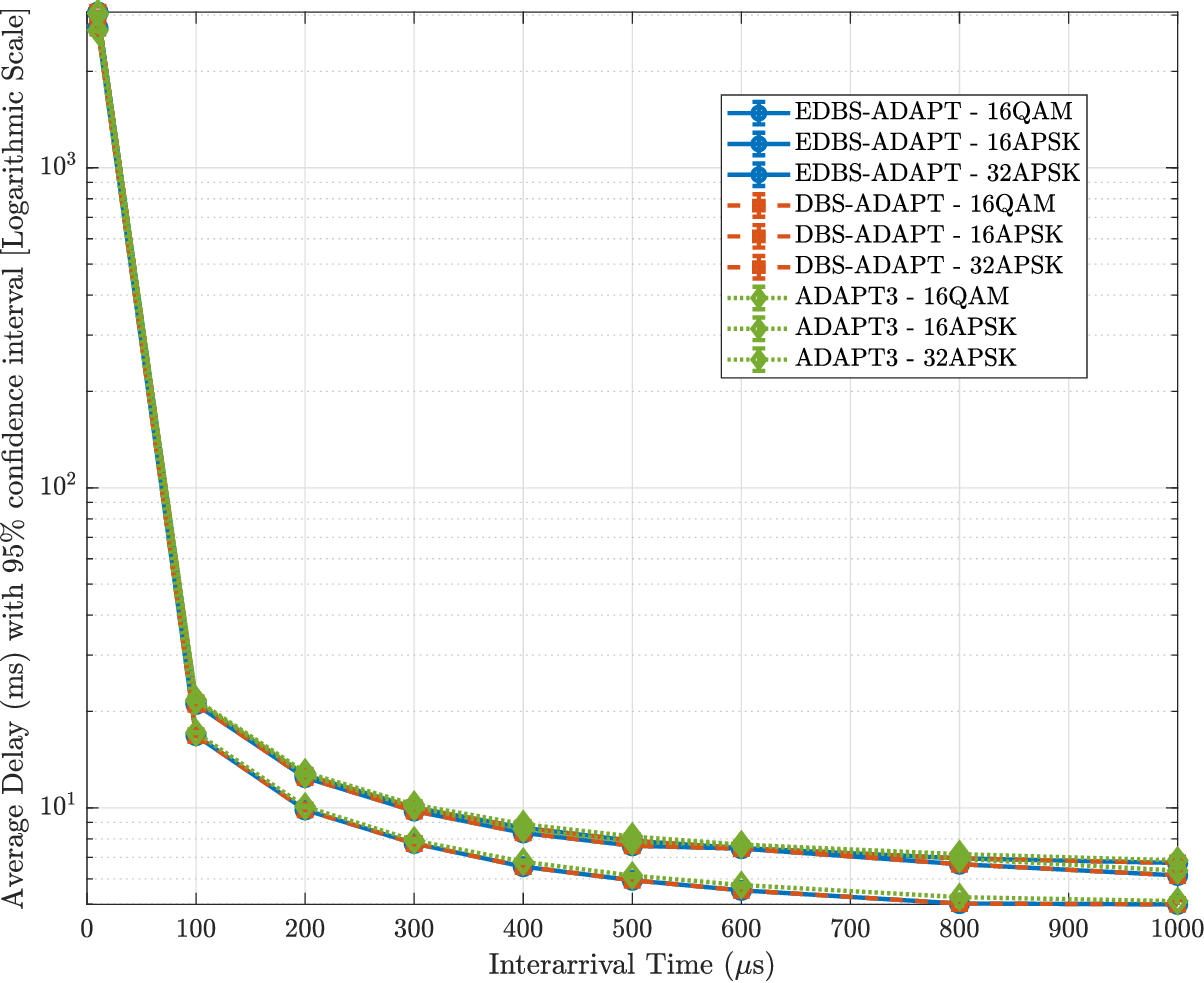}}
    \caption{EDBS-ADAPT 14 nodes case with varying inter-arrival time.}
    \label{fig:18}
\end{figure}

Even under high traffic conditions, where the interarrival time is as low as 10 $\mu\text{s}$, EDBS-ADAPT and DBS-ADAPT maintains better performance. In this worst-case scenario, EDBS-ADAPT and DBS-ADAPT achieves an average throughput of 0.65 Gbps, compared to 0.63 Gbps for ADAPT-3, representing a 3\% improvement. These results demonstrates EDBS-ADAPT and DBS-ADAPT superior ability to handle varying traffic loads, particularly in maintaining higher throughput under both light and heavy conditions.

\begin{figure}[h]
    \centering
    \subfigure[Average throughput with BPSK, QPSK and 8PSK MCS]{\label{fig:varying_nodes_thr_1} \includegraphics[width=0.38\textwidth]{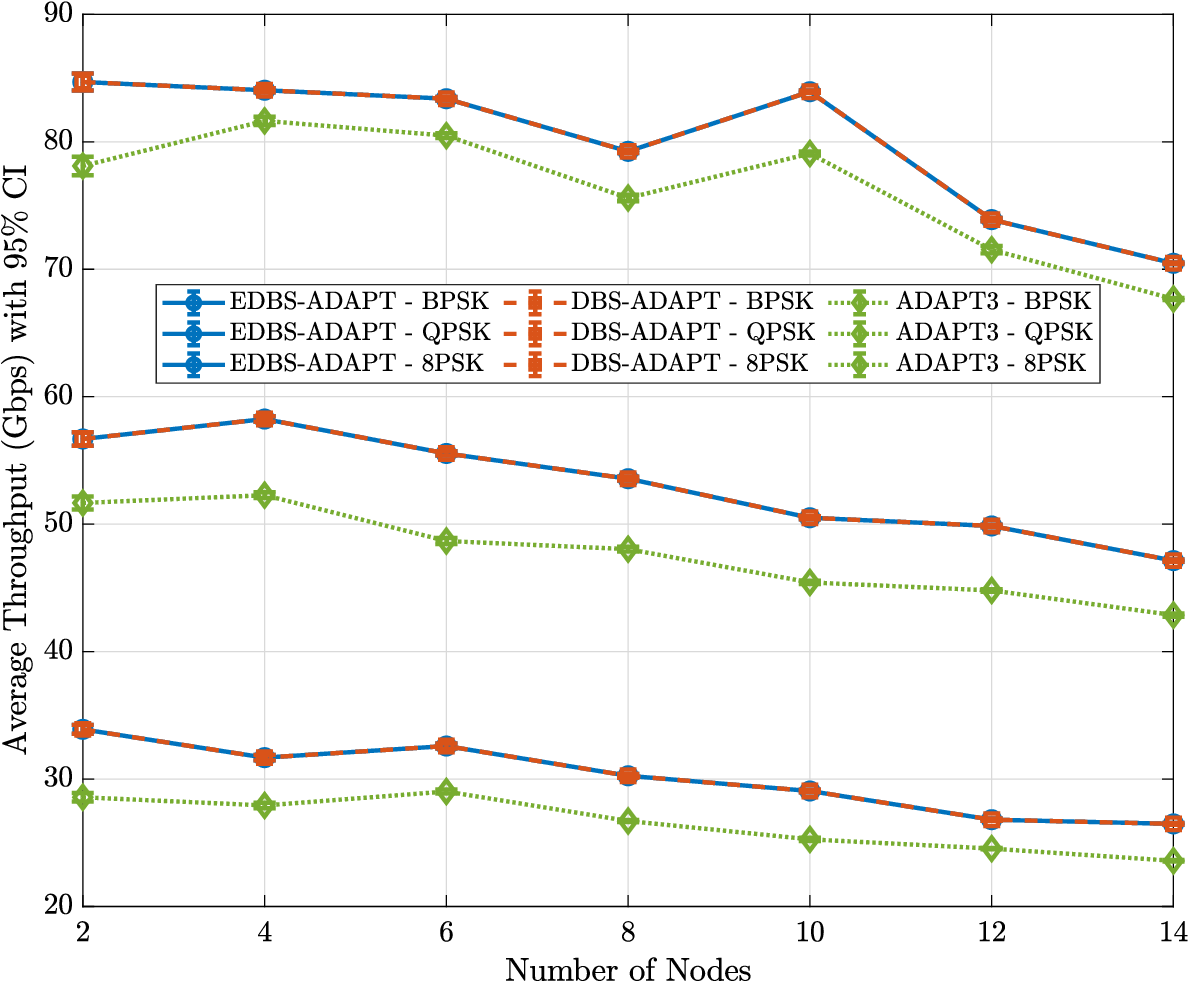}}
    \hfill
    \subfigure[Average delay with BPSK, QPSK and 8PSK MCS]{\label{fig:varying_nodes_ad_1} \includegraphics[width=0.38\textwidth]{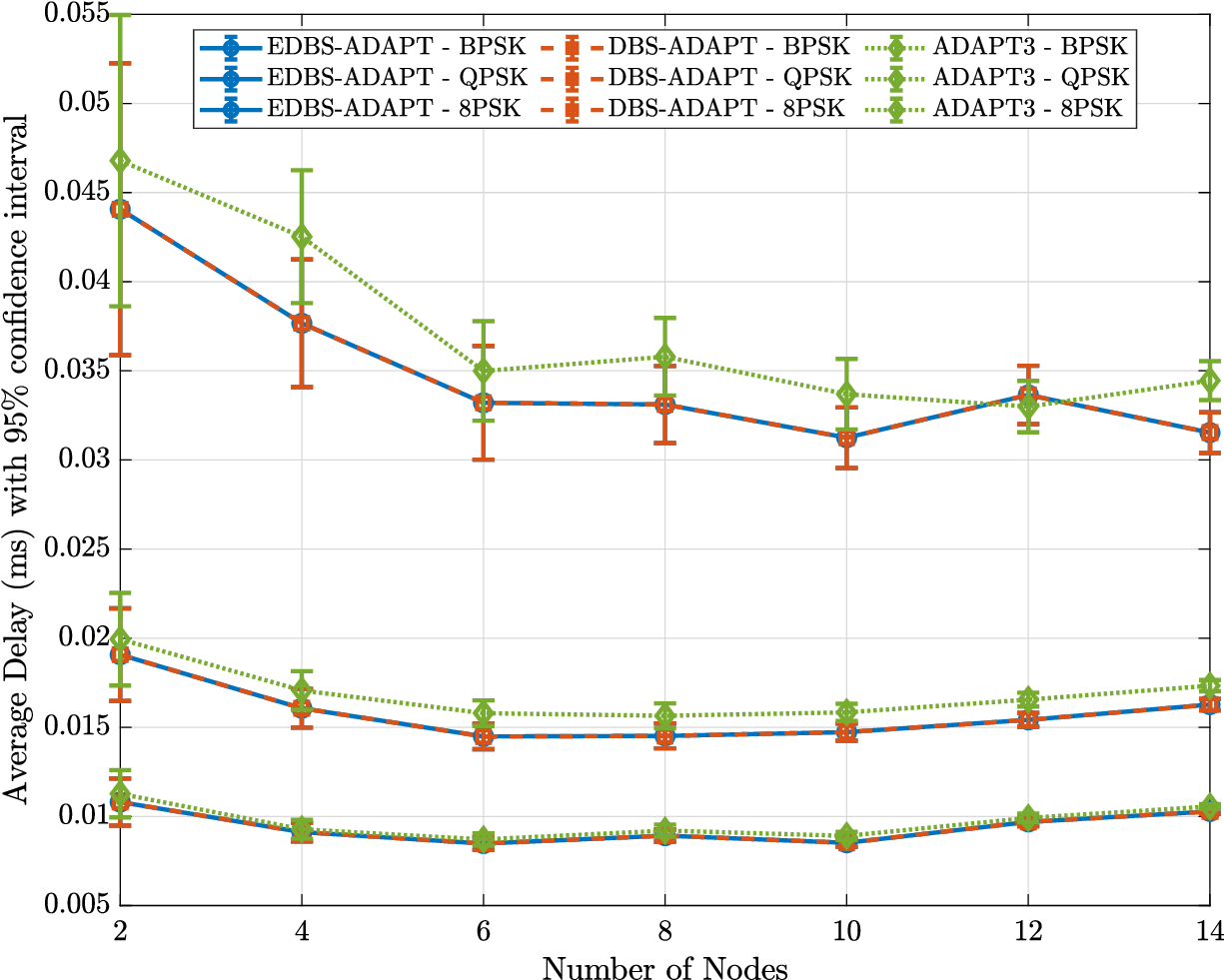}}
    \caption{EDBS-ADAPT with varying nodes case and interarrival time 500 $\mu\text{s}$}
    \label{fig:edbs_var_node_1}
\end{figure}

The average delay results for EDBS-ADAPT and DBS-ADAPT, is shown in Figure 5(b), indicate a consistent reduction compared to the ADAPT-3 THz MAC protocol when using the BPSK modulation scheme. Under low traffic conditions with an interarrival time of 1000 $\mu\text{s}$, EDBS-ADAPT and DBS-ADAPT achieves an average delay of 26 ms, compared to 29 ms for ADAPT-3, a reduction of approximately 10\%. With an interarrival time of 500 $\mu\text{s}$, EDBS-ADAPT and DBS-ADAPT records a delay of 29 ms, while ADAPT-3 has 32 ms, showing a 10\% improvement. Even under high traffic conditions (interarrival time of 10 $\mu\text{s}$), EDBS-ADAPT and DBS-ADAPT reduces average delay by 9\% compared to ADAPT-3.

This trend continues across higher modulation schemes, as illustrated in Figures 6(a) and 6(b). In both high and low traffic scenarios, EDBS-ADAPT and DBS-ADAPT consistently outperforms ADAPT-3 in terms of average throughput and delay. These results demonstrate EDBS-ADAPT and DBS-ADAPT effectiveness in improving communication efficiency, particularly in reducing latency and maintaining performance across a range of traffic loads and modulation settings.

\begin{figure}[h]
    \centering
    \subfigure[Average throughput with 16QAM, 16APSK, and 32APSK MCS.]{\label{fig:varying_nodes_thr_2a} \includegraphics[width=0.38\textwidth]{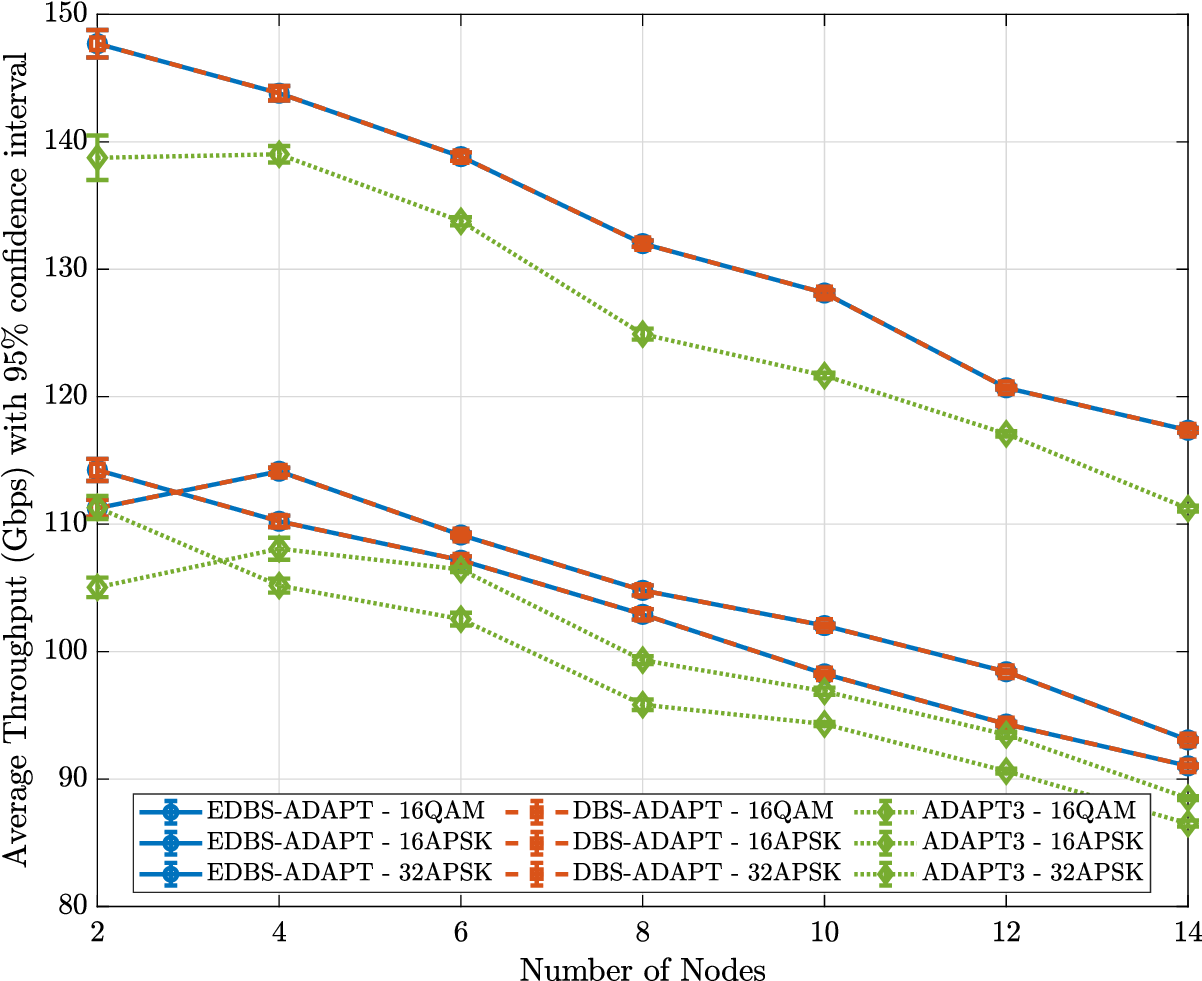}}
    \hfill
    \subfigure[Average delay with 16QAM, 16APSK, and 32APSK MCS.]{\label{fig:varying_nodes_thr_2b} \includegraphics[width=0.38\textwidth]{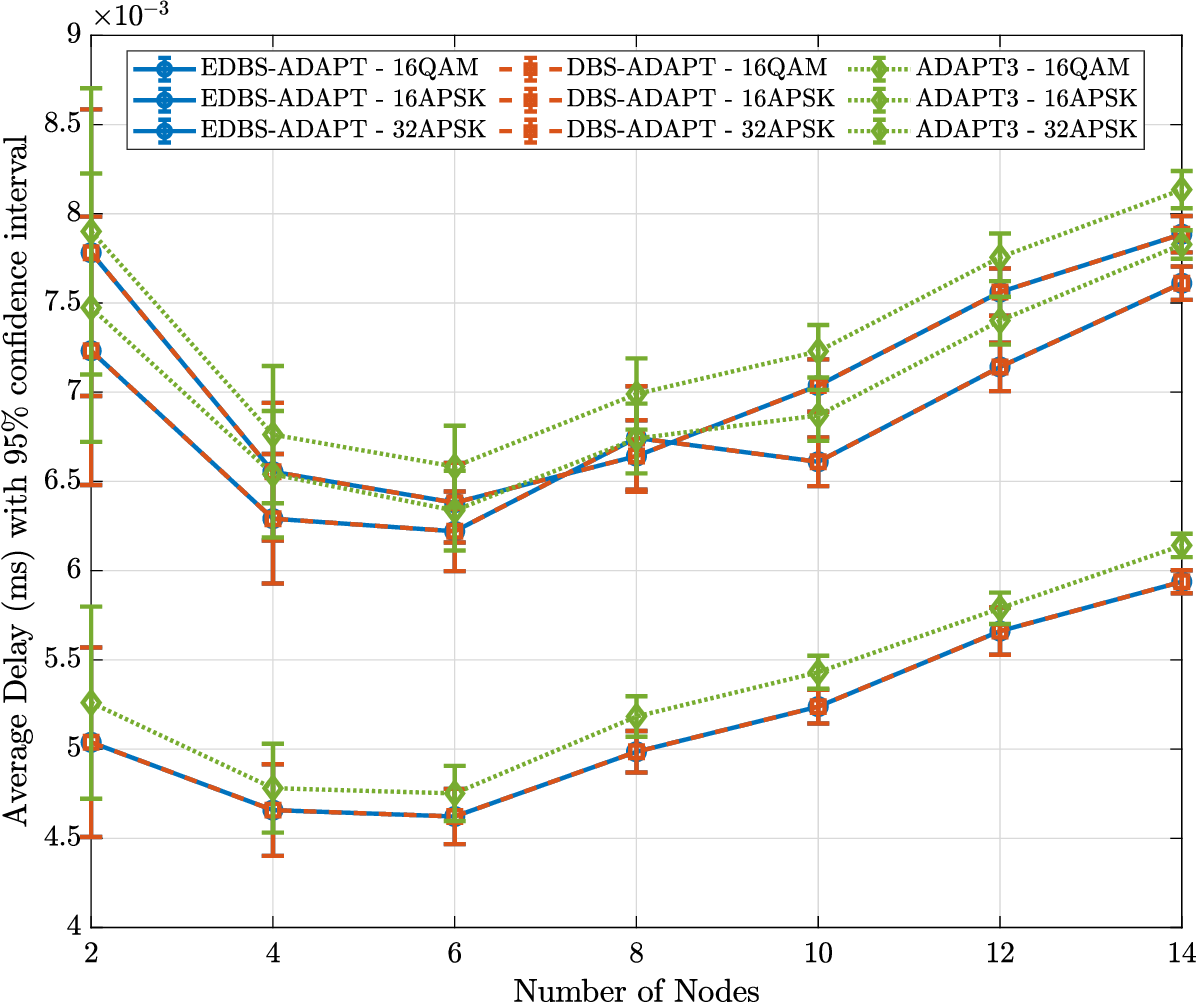}}
    \caption{EDBS-ADAPT with varying nodes case and interarrival time 500 $\mu\text{s}$.}
    \label{fig:edbs_var_node_2}
\end{figure}

EDBS-ADAPT and DBS-ADAPT consistently outperforms ADAPT-3 across all evaluated modulation schemes, BPSK, QPSK, 8-PSK, 16-QAM, 16-APSK, and 32-APSK, showing both higher average throughput and lower average delay. This performance holds across varying traffic loads, from low to high. The consistent improvement demonstrates that EDBS-ADAPT and DBS-ADAPT effectively addresses the unique challenges of different modulation levels: providing robustness for lower-order schemes and maintaining precision and efficiency for higher-order schemes. This adaptability makes EDBS-ADAPT and DBS-ADAPT well-suited for diverse communication requirements in THz based data center environments.
\paragraph{\textbf{Beamwidth switching overhead with varying inter-arrival times and modulation schemes:}} Table 5 presents the beamwidth switching overhead for EDBS-ADAPT, DBS-ADAPT, and ADAPT-3 under a fixed 14-node setup with varying interarrival times. ADAPT-3 overhead is 0 for all the scenarios as it uses the fixed beamwidth. While, EDBS-ADAPT beamwidth switching overhead is minimised by upto 95\% for both the AP and client nodes for all the case as compared to DBS-ADAPT which have high beamwidth switching overhead.
\begin{table*}[ht]

    \centering
    \small 
    \caption{Beamwidth-Switching Overhead 14 nodes case with varying inter-arrival time}
    \label{tab:beamwidth_switching}
        \begin{tabularx}{\textwidth}{|p{3cm}|X|X|X|X|X|}

        \toprule
      \textbf{Interarrival Time $\mu\text{s}$} 
          & \textbf{DBS-ADAPT client node overhead} 	& \textbf{EDBS-ADAPT client node overhead per node} & \textbf{EDBS-ADAPT AP node overhead}  & \textbf{EDBS-ADAPT AP node overhead per node}	\\
        \midrule
        10   & 210	& 1	& 210 	& 1   \\
        100   & 90 	&  1  	 	& 90 	&  1     \\
        200   & 47	&  1  	 	& 47 	&  1     \\
        300  & 31	&  1   	 	& 31 	&  1      \\
        400  & 24	&  1  	 	& 24 	&  1      \\
        500  & 19 	&  1  	 	& 19 	&  1    \\
        600  & 16 	&  1  	 	& 16 	&  1     \\
        800  & 12 	&  1   	 	& 12 	&  1     \\
        1000 & 10 	&  1  	 	& 10 	&  1    \\
        \bottomrule
    \end{tabularx}
    \vspace{6pt} 

    {\footnotesize
    \textbf{Note:}
    \begin{itemize}
          \item \textbf{DBS-ADAPT:} DBS-ADAPT overhead is for both AP and client node and it is the average adjustment per node.
      \item \textbf{EDBS-ADAPT:} Overhead is for per client node.
      \item \textbf{ADAPT-3:} AP overhead = 0, Client overhead = 0 (constant for all interarrival times).
    \end{itemize}
    }
\end{table*}

\subsubsection{EDBS-ADAPT, and DBS-ADAPT with varying number of nodes and modulation schemes}
The performance of the EDBS-ADAPT and DBS-ADAPT THz MAC protocols is evaluated by varying the number of nodes and applying different modulation and coding schemes, as detailed in Table 3. The number of nodes ranges from 2 to 14, while six MCS configurations are considered. For this evaluation, the interarrival time is fixed at 500 $\mu\text{s}$, corresponding to a traffic load of 20 packets transmitted by each client node. The resulting average throughput and average delay for each protocol, across different node counts, are presented in Figures 7 and 8.

\paragraph{\textbf{Average Throughput and delay with varying nodes and modulation schemes}}
In the 2-node scenario using the lowest modulation scheme (BPSK), both EDBS-ADAPT and DBS-ADAPT achieved an average throughput of approximately 34 Gbps, while ADAPT-3 reached only 28 Gbps, as illustrated in Fig. 7(a). This represents a 21\% increase in throughput for EDBS-ADAPT and DBS-ADAPT compared to ADAPT-3. While the average delay reduced by 6\% compared to ADAPT-3 as shown in Fig.7(b). Similarly for the 10 nodes scenario, Additionally, these protocols reduced the average delay by about 5\% as shown in Fig. 7(b).

Similarly, in the 10-node scenario, EDBS-ADAPT and DBS-ADAPT demonstrated a 15\% improvement in average throughput and a 7\% reduction in average delay compared to ADAPT-3.

In the 2-node scenario using the highest modulation scheme (32-APSK), EDBS-ADAPT and DBS-ADAPT achieved an average throughput of approximately 148 Gbps, while ADAPT-3 reached 137 Gbps, as shown in Fig. 8(a). This corresponds to a 8\% cincrease in throughput for EDBS-ADAPT and DBS-ADAPT compared to ADAPT-3. Additionally, the average delay was reduced by about 4\% as illustrated in Fig.8(b).

A similar trend was observed in the 10-node scenario, where EDBS-ADAPT and DBS-ADAPT maintained a 5\% higher average throughput and a 4\% lower average delay compared to ADAPT-3.

\paragraph{\textbf{Beamwidth switching overhead for varying nodes and modulation schemes:}} Table 6 compares the beamwidth switching overhead of EDBS-ADAPT, DBS-ADAPT, and ADAPT-3 in a fixed 14-node setup with varying interarrival times. ADAPT-3 incurs no overhead due to its fixed beamwidth, while EDBS-ADAPT reduces switching overhead by up to 94.74\% for both AP and client nodes compared to the higher overhead of DBS-ADAPT.

\begin{table*}[ht]
    \small 
    \centering
    \caption{Beamwidth-Switching Overhead with varying nodes case and interarrival time 500 $\mu\text{s}$}
    \label{tab:beamwidth_overhead}
    \begin{tabular}{@{}l  c c p{0.35\textwidth}@{}}
        \toprule
        \textbf{Name} & \textbf{AP node overhead} & \textbf{Client node overhead} & \textbf{Details} \\
        \midrule
        \textbf{EDBS-ADAPT} & 1  & 1 (Per Client Node) &
        Each client node and the AP node adjust their beamwidth only once per node.
    \\
        \textbf{DBS-ADAPT} & 19 (Avg adjustment per node) & 19 (Avg adjustment per node) &
        Both the AP and each client node adjusted their beamwidth multiple times. 
        Here 19 is the \emph{average} number of adjustments per node.  \\
        \textbf{ADAPT-3 \cite{morales2021adapt}} & 0 & 0 &
       Both AP and client nodes use a fixed beamwidth. \\
        \bottomrule
    \end{tabular}
\end{table*}

Figures 7 and 8 show that EDBS-ADAPT and DBS-ADAPT consistently achieve higher average throughput and lower average delay than ADAPT-3 across all tested modulation schemes (BPSK to 32-APSK) and node counts from 2 to 14.

\begin{figure}
	\centering
	\includegraphics[width=.9\columnwidth]{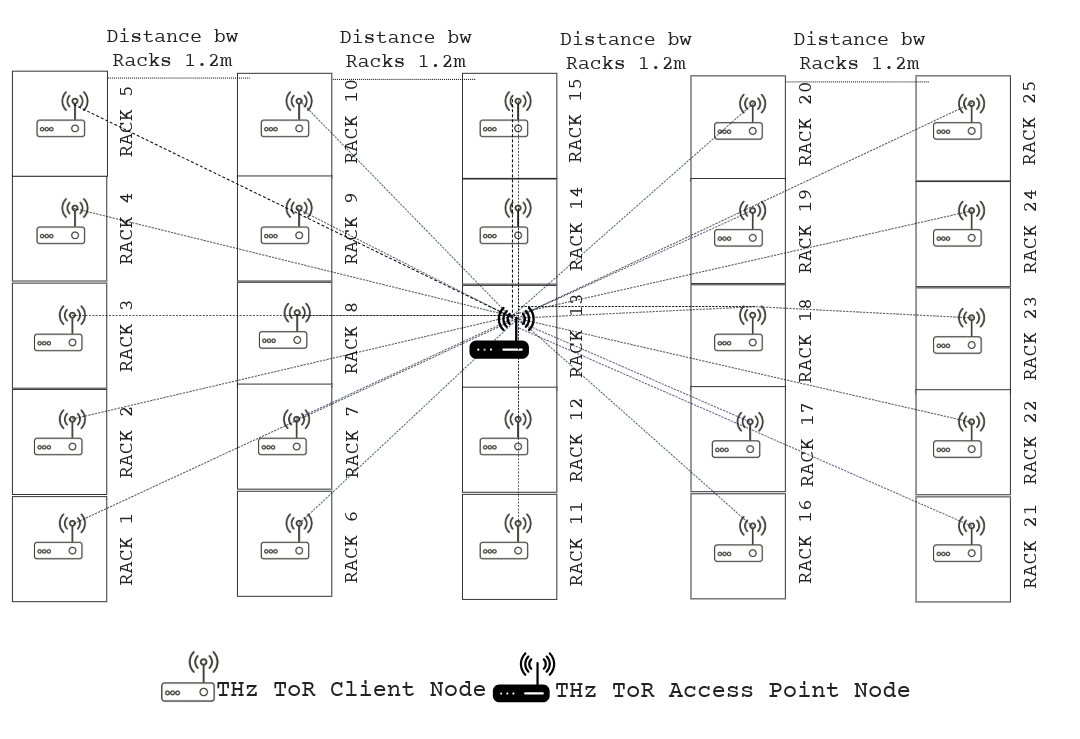}
	\caption{Topology for 25-node scalability scenario.}
	\label{FIG:1}
\end{figure}
\subsubsection{Scalability Analysis} The performance of the EDBS-ADAPT and DBS-ADAPT THz MAC protocols is evaluated by increasing the number of nodes to 25 as shown in Fig. 9. In this scenario, there are 24 client nodes and 1 AP node. In this evaluation, the interarrival time is fixed at 500 $\mu\text{s}$. This evaluation is conducted to determine if the expected throughput gains are sustained at a larger scale.

\begin{figure}[h]
    \centering
    \subfigure[Average throughput with 16QAM, 16APSK, and 32APSK MCS.]{\label{fig:varying_nodes_thr_2a} \includegraphics[width=0.38\textwidth]{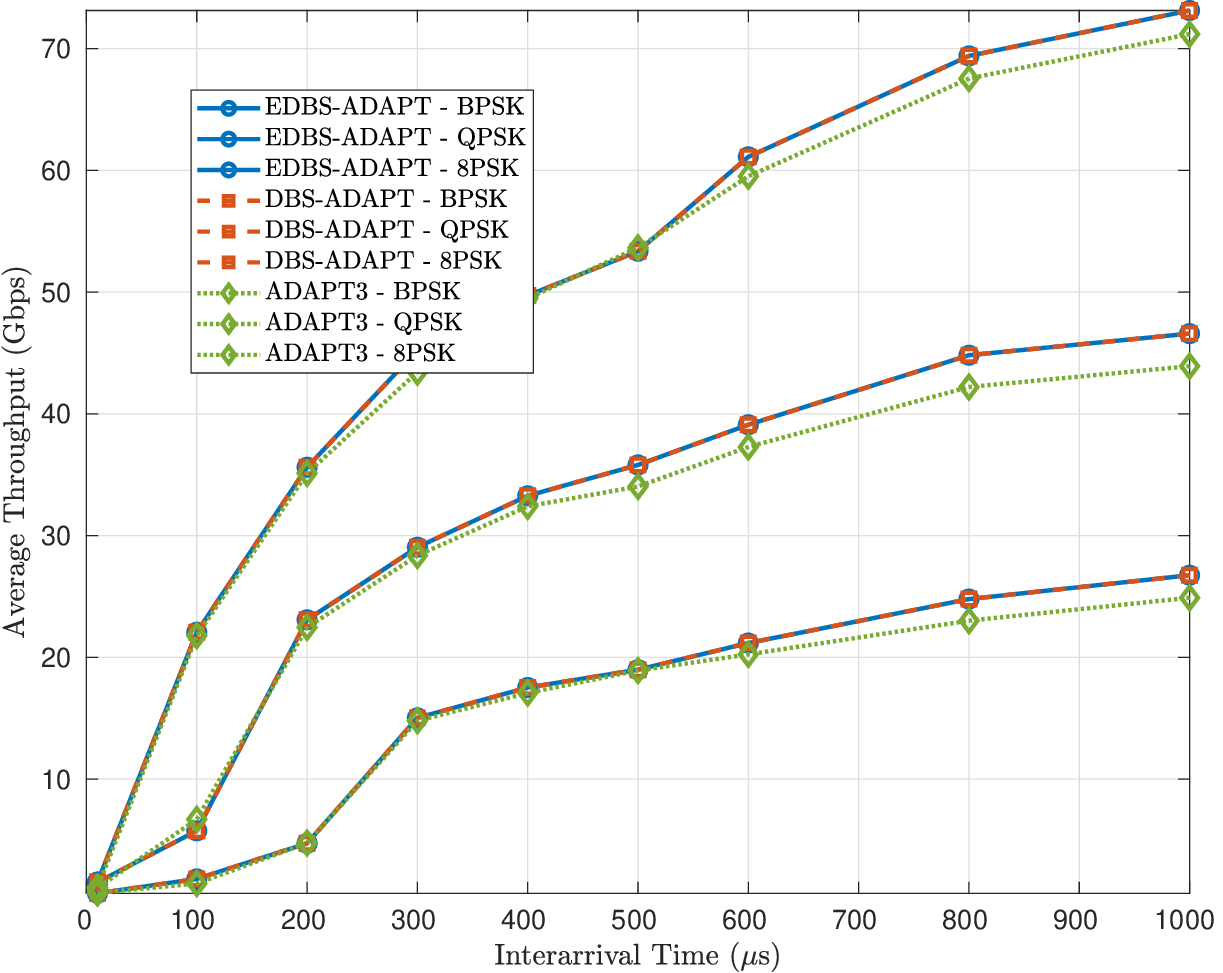}}
    \hfill
    \subfigure[Average delay with 16QAM, 16APSK, and 32APSK MCS.]{\label{fig:varying_nodes_thr_2b} \includegraphics[width=0.38\textwidth]{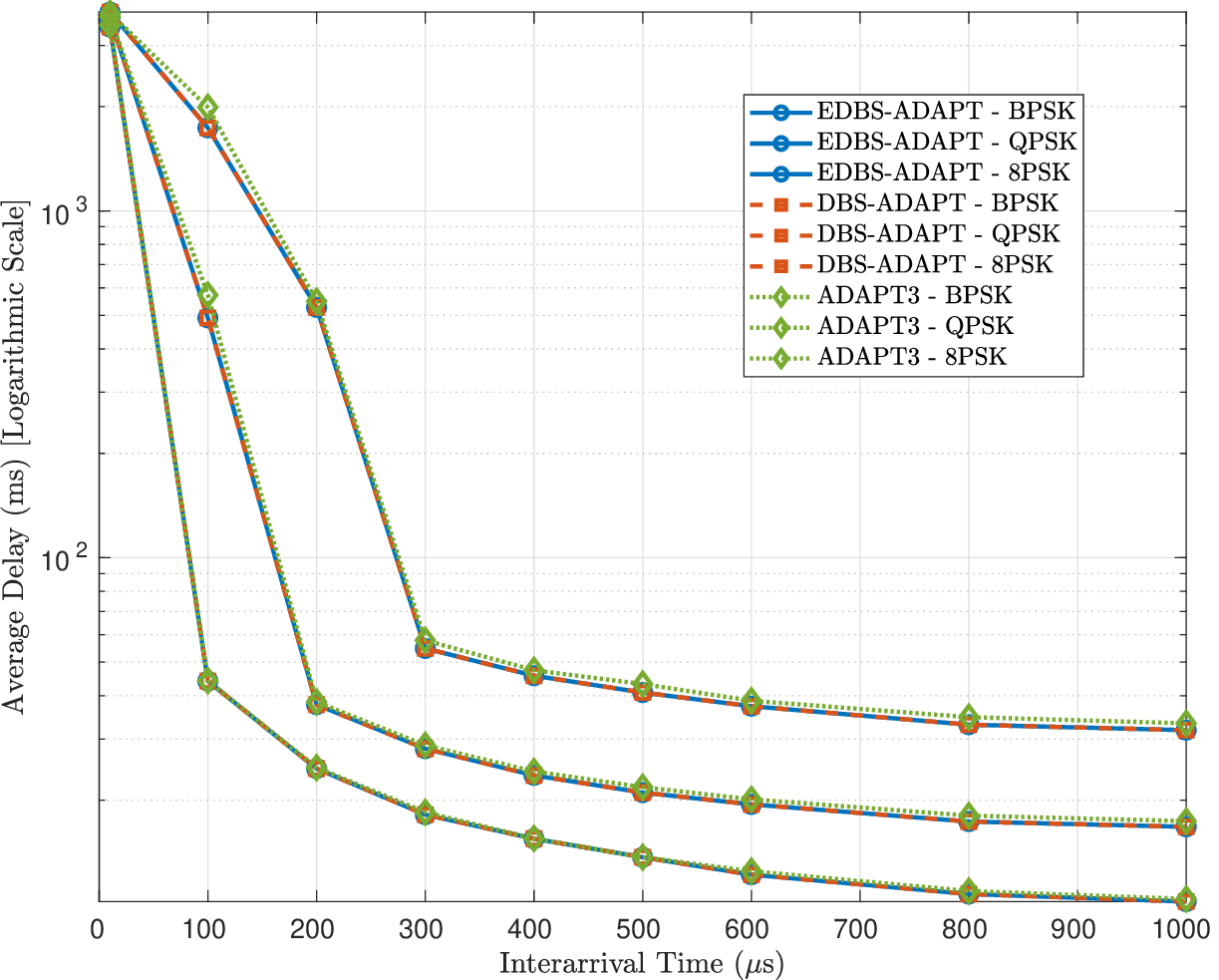}}
    \caption{EDBS-ADAPT, DBS-ADAPT and ADAPT-3 performance in the 25-node scalability scenario.}
    \label{fig:edbs_var_node_2}
\end{figure}

The resulting average throughput and average delay for each protocol are presented in Figures 10(a) and 10(b). In this scenario, EDBS-ADAPT and DBS-ADAPT maintained a higher average throughput and a lower average delay compared to ADAPT-3. The results shows the advantages of the proposed protocol even in the case of scalability when the number of node increases.

\subsubsection{Discussion on Improved Performance}
The increased average throughput of EDBS-ADAPT and DBS-ADAPT is primarily due to its dynamic beamwidth selection during data transmission. By using a wider beamwidth, the number of sectors decreases, allowing the antenna to rotate faster and create more frequent communication opportunities between the AP and client nodes. This faster sector scanning reduces waiting times and delays, leading to higher throughput. The key factor driving this improvement is the antenna’s turning speed, which depends on the beamwidth and sector timing. Since turning speed is directly proportional to beamwidth (as shown in Fig. 4), wider beams enable quicker coverage of the communication area, positively impacting both throughput and delay.

Higher-order modulation schemes achieve greater average throughput compared to lower-order schemes, mainly due to their higher physical layer data rates. For example, 32-APSK yields the highest throughput, followed by 16-APSK, 16-QAM, 8-PSK, QPSK, and BPSK. This is because higher-order modulations encode more bits per symbol, allowing more data to be transmitted within the same time frame. Conversely, lower-order schemes like BPSK carry fewer bits, resulting in lower throughput despite their robustness.

ADAPT-3 achieves lower average throughput compared to DBS-ADAPT and EDBS-ADAPT due to its use of a fixed beamwidth. Fixed beamwidth prevents nodes from adapting to varying distances, wider beams are better for nearby nodes to reduce delay, while narrower beams are ideal for distant nodes to maintain connectivity. This lack of adaptability leads to reduced throughput and increased delay, which is the primary cause of ADAPT-3's performance decline in the THz MAC protocol.

Overall, our detailed evaluations under varying interarrival time and varying nodes clearly shows that EDBS-ADAPT achieves higher throughput and less delay while EDBS-ADAPT achieves identical throughput and delay to DBS-ADAPT as it follows the same core mechanism as DBS-ADAPT. However, the advantage of our EDBS-ADAPT lies in substantially reduced antenna beamwidth switching overhead, while preserving the throughput and delay advantages of the original protocol.

\subsection{Simulation Results with Bursty Traffic}\label{sec:periodicityyu}
Data center traffic is often bursty, with periods of intense activity followed by idle intervals \cite{285196}. This burstiness impacts the performance of THz MAC protocols, which depend on precise timing and coordination between nodes. To simulate such traffic, the NS-3 Terasim simulator was modified to use a log-normal distribution in the traffic generator, accurately modeling the bursty behavior observed in data centers. The traffic generator class was specifically extended to produce packets based on this distribution \cite{ns3lognormal}.

In our simulation, the log-normal distribution parameters were set to (\(\mu\)) $ = 5 $ and (\(\sigma\)) $ = 1 $, reflecting typical bursty traffic patterns observed in data center studies \cite{285196}. These values were chosen to capture moderate average interarrival times (\(\mu\)) and high variability (\(\sigma\)), effectively modeling the unpredictable nature of data center traffic with bursts of high packet arrivals followed by idle periods \cite{285196}. Additionally, in the bursty traffic scenario we evaluated with 15 nodes scenario, in which there are 14 client nodes.

Using the lowest modulation scheme (BPSK), EDBS-ADAPT and DBS-ADAPT consistently outperformed ADAPT-3 in both throughput and delay. In the 2-node scenario, they achieved approximately 33 Gbps compared to ADAPT-3's 27 Gbps, a 22\% increase. For 10 nodes, the throughput gain was around 22\% (28 Gbps vs. 23 Gbps). In terms of delay, EDBS-ADAPT and DBS-ADAPT reduced average latency by 9\% in the 2-node case and by 10\% in the 10-node case, as shown in Figures 11(a) and 11(b).

\begin{figure}[h]
    \centering
    \subfigure[Average throughput with BPSK, QPSK and 8PSK MCS.]{\label{fig:avg_thr_burs_1} \includegraphics[width=0.38\textwidth]{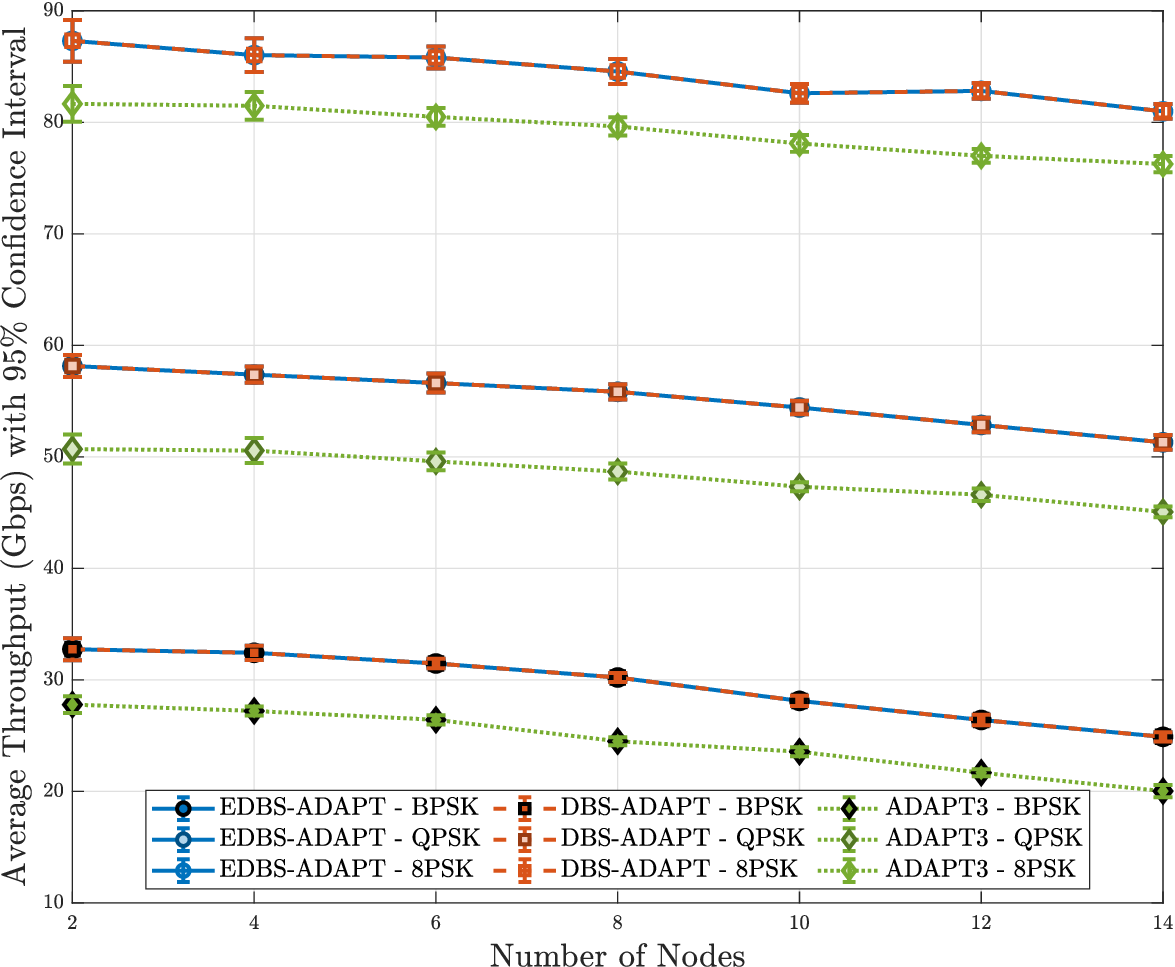}}
    \hfill
    \subfigure[Average delay with BPSK, QPSK and 8PSK MCS.]{\label{fig:avg_ursyu_2} \includegraphics[width=0.38\textwidth]{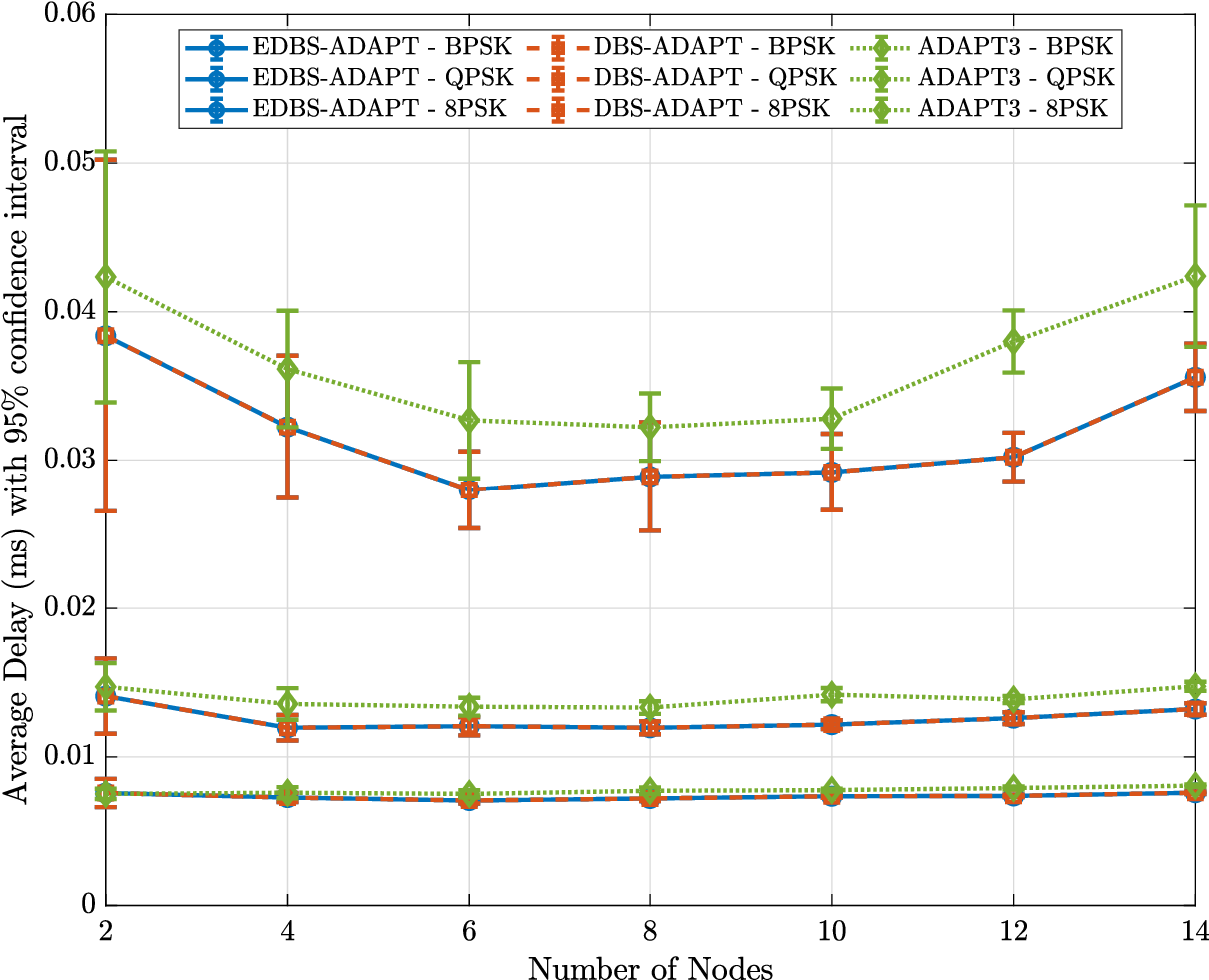}}
    \caption{EDBS-ADAPT, DBS-ADAPT and ADAPT-3 with varying nodes case and bursty traffic.}
    \label{fig:edbs_bursty_1}
\end{figure}

\begin{table*}[ht] 
    \small 

\centering 

\begin{tabularx}{\columnwidth}{|X|X|X|} 

\hline 

\textbf{Description} & \textbf{Value} \\ 

\hline 

Traffic Model & Log normal (bursty) \cite{ns3lognormal} \\ 

\hline 

Simulation time for each run &  0.01 seconds \\ 

\hline 

Simulation Runs &  20 \\ 

\hline

Log-Normal (\(\mu\))  & 5 \\ 

\hline 

Log-Normal (\(\sigma\)) & 1 \cite{285196} \\ 

\hline 

 \end{tabularx} 
\caption{Summary of Log-normal bursty-traffic Parameters used in the simulation \label{tab:table1} }
\end{table*}

\begin{figure}[h]
    \centering
    \subfigure[Average throughput with 16QAM, 16APSK, and 32APSK MCS.]{\label{fig:avg_thr_burs454_2} \includegraphics[width=0.38\textwidth]{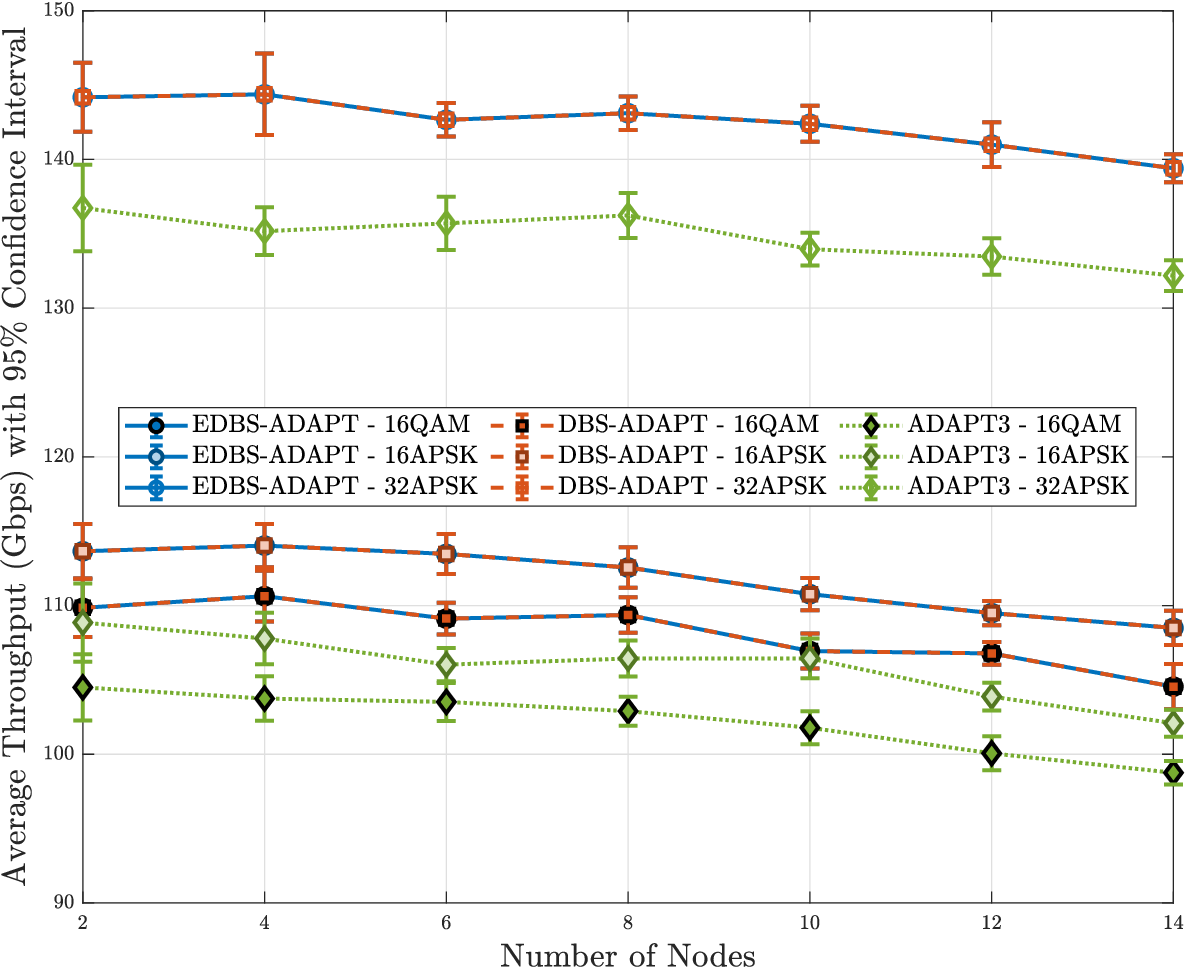}}
    \hfill
    \subfigure[Average delay with 16QAM, 16APSK, and 32APSK MCS.]{\label{fig:avg_pd_burs_2} \includegraphics[width=0.38\textwidth]{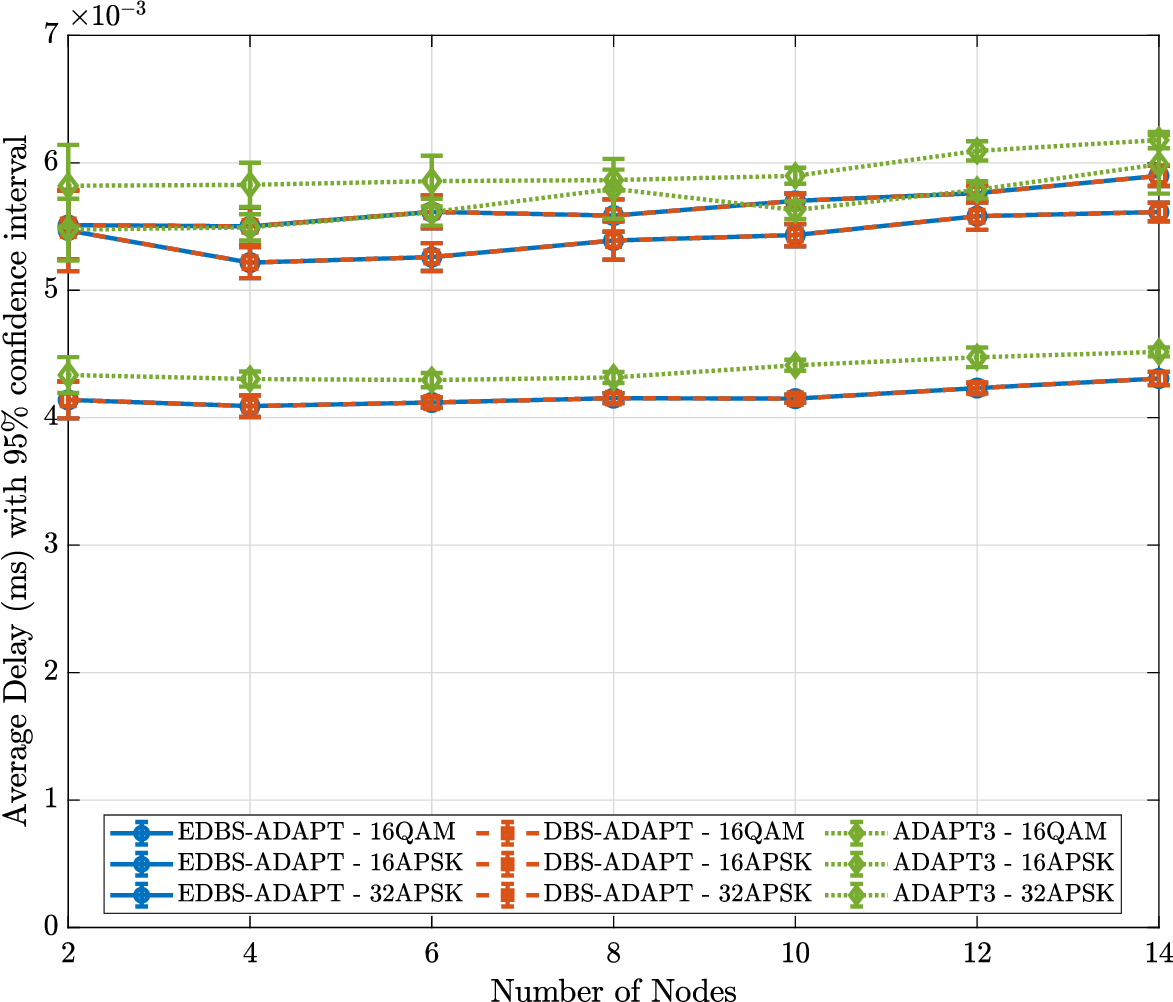}}
    \caption{EDBS-ADAPT, DBS-ADAPT and ADAPT-3 with varying nodes case and bursty traffic.}
    \label{fig:edbs_bursty_2}
\end{figure}
At the highest modulation scheme (32-APSK), EDBS-ADAPT and DBS-ADAPT outperformed ADAPT-3 in both throughput and delay. In the 2-node scenario, they achieved approximately 144 Gbps, about 6\% higher than ADAPT-3's 136 Gbps, with an 4\% reduction in average delay, as shown in Figures 12(a) and 12(b). For 10 nodes, throughput improved by 7\%, and delay was reduced by 6\% compared to ADAPT-3.

In scenarios with varying numbers of nodes, average throughput remains stable due to the heavy-tailed nature of the log-normal distribution, which generates bursts of high activity followed by extended silent periods. This allows nodes to transmit packets quickly and then remain silent for a while. As the number of nodes increases, the likelihood of all nodes transmitting simultaneously stays low, reducing channel contention and maintaining throughput stability.

Figures 11 and 12 show that EDBS-ADAPT and DBS-ADAPT consistently achieve higher throughput and lower delay than ADAPT-3 across all modulation schemes (BPSK to 32-APSK) and node counts from 2 to 14. This performance improvement is attributed to their dynamic beamwidth adjustment, unlike ADAPT-3's fixed beamwidth approach.

\section{Conclusion}\label{sec:conclusion} 
This paper proposes DBS-ADAPT, a dynamic beamwidth selection-based THz MAC protocol for wireless data centers, along with its enhanced version, EDBS-ADAPT. Both protocols overcome the limitations of fixed beamwidth by dynamically selecting the optimal beamwidth based on node distance, which improves throughput and reduces delay. By exchanging node position information, the protocols ensure that the chosen beamwidth maintains communication range without compromise, enhancing overall protocol efficiency. The performance of DBS-ADAPT and EDBS-ADAPT was evaluated through simulations using the Terasim (NS3) THz simulator with realistic data center parameters. The results demonstrate that both protocols achieve higher average throughput and lower delay compared to the baseline ADAPT protocol across various traffic loads, node counts, and modulation schemes. Additionally, EDBS-ADAPT reduces antenna overhead compared to DBS-ADAPT.

Future work will focus on scaling the evaluation to larger data center environments with multiple APs to test the protocols' scalability and robustness. Additionally, we plan to extend our approach by incorporating an adaptive, distance aware power allocation strategy to improve energy efficiency.
\bibliographystyle{elsarticle-num}

\bibliography{cas-refs}


\end{document}